\documentclass[reprint, superscriptaddress, amsmath,prl,amssymb, aps, floatfix]{revtex4-2}

\makeatletter
\newsavebox{\@brx}
\newcommand{\llangle}[1][]{\savebox{\@brx}{\(\m@th{#1\langle}\)}%
  \mathopen{\copy\@brx\kern-0.5\wd\@brx\usebox{\@brx}}}
\newcommand{\rrangle}[1][]{\savebox{\@brx}{\(\m@th{#1\rangle}\)}%
  \mathclose{\copy\@brx\kern-0.5\wd\@brx\usebox{\@brx}}}
\usepackage{lipsum}
\usepackage{graphicx}% Include figure files
\usepackage{dcolumn}% Align table columns on decimal point
\usepackage{bm}% bold math
\usepackage{xcolor}
\usepackage{xspace}
\usepackage[colorlinks,allcolors=blue]{hyperref}
\usepackage[makeroom]{cancel}
\usepackage{float}
\usepackage{siunitx}
\usepackage{color}
\definecolor{dgreen}{rgb}{0,0.7,0}
\def\bluerev#1{{\color{black} #1}}
\def\redw#1{{\color{black} #1}}

\def\greenw#1{{\color{black} #1}}
\def\bluew#1{{\color{black} #1}}
\def\ps#1{{\color{black} #1}}
\newcommand{\beq}{\begin{equation}}
\newcommand{\eeq}{\end{equation}}
\newcommand{\bea}{\begin{eqnarray}}
\newcommand{\eea}{\end{eqnarray}}

\usepackage{subfigure}
\usepackage{booktabs} 
\usepackage{natbib}
\usepackage[normalem]{ulem}

\begin{document}
%\title{Thermodynamic limit to positional information}
\title{Limits to positional information in boundary-driven systems}

\author{Prashant Singh}
\email{prashantsinghramitay@gmail.com}
\author{Karel Proesmans}
\email{karel.proesmans@nbi.ku.dk}
%\address{International Centre for Theoretical Sciences, Tata Institute of Fundamental ~Research, Bengaluru 560089, India $^1$}
\address{Niels Bohr International Academy, Niels Bohr Institute,
University of Copenhagen, Blegdamsvej 17, 2100 Copenhagen, Denmark}
%\ead{karel.proesmans@nbi.ku.dk}
%\ead{prashant.singh@icts.res.in, saikat.santra@icts.res.in}
%\ead{anupam.kundu@icts.res.in}
\vspace{10pt}

\begin{abstract}
Chemical gradients can be used by a particle to determine its position. This \textit{positional information} is of crucial importance, for example in developmental biology \redw{in the formation of patterns in an embryo}. The central goal of this paper is to study the fundamental physical limits on how much positional information can be stored inside a system. To achieve this, we study positional information for boundary-driven systems, and derive, in the near-equilibrium regime, a universal expression involving only the chemical potential and density gradients of the system. We also conjecture that this expression serves as an upper bound on the positional information of boundary driven systems beyond linear response. To support this claim, we test it on a broad range of solvable boundary-driven systems. 
\end{abstract}
\maketitle

\noindent
\textit{Introduction:} Nature endows biological matter with the astounding ability to self-organize fascinating spatio-temporal patterns that pervade across several length scales \cite{bookMeinhardt}. For instance, one often sees insects with their bodies divided into segments of repetitive patterns or birds and animals with unique patterns of spots and stripes. One classic mechanism of pattern formation in reaction-diffusion systems is Turing pattern \cite{Turingcite,falasco2018information,rana2020precision}. While naively, one expects diffusion to generate uniform concentration of chemicals, Turing showed that two diffusing chemical species with distinct diffusion coefficients and activation-inhibition interplay can, under suitable condition, spontaneously break the symmetry of homogeneous concentrations and generate recurring structures such as stripes, spots, or even more complex patterns.
 
\begin{figure}[t]
\centering
  \includegraphics[scale=0.4]{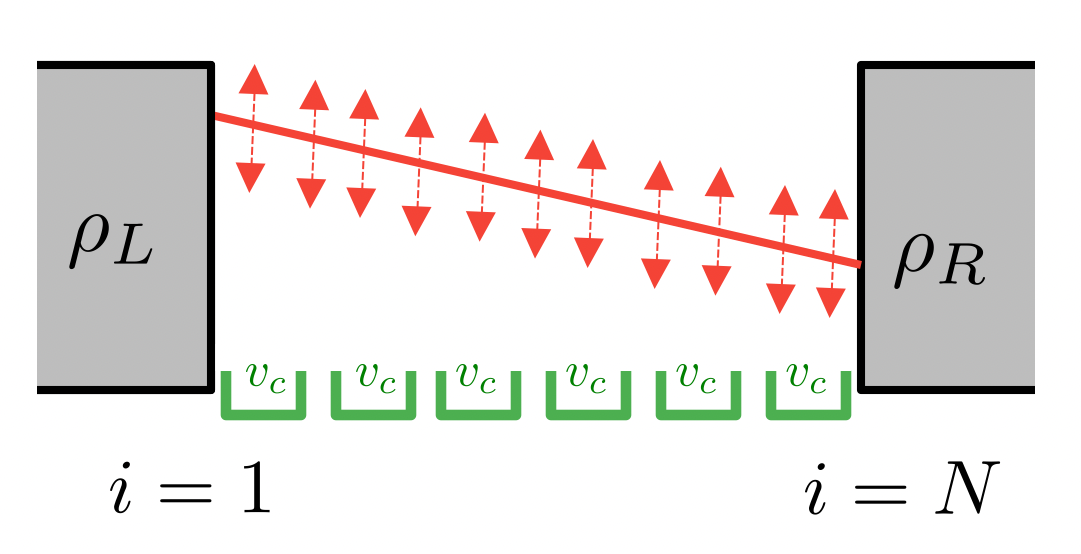}
  \caption{\redw{Lattice sites depicted in green are in contact with two distinct reservoirs characterized by average densities $\rho _L$ and $\rho _R$. Due to this coupling, the system attains a steady-state with average density profile shown by the red solid line. However, due to noise, there are fluctuations around this mean value which are indicated by dashed arrows. Positional information of a lattice site is obtained \ps{by measuring the local particle density inside a small volume $v_c$} and employing the formula in Eq.~\eqref{avg-information-new}.}}
%They can accommodate particles whose 
%  Cells inside an embryo are represented by lattice sites (depicted in green) that spread across $x=0$ and $x=1$ ($x$ being some rescaled coordinate). The ends are in contact with two distinct reservoirs characterized by average densities $\rho _L$ and $\rho _R$. The steady-state average density profile inside the system is shown by the red solid line. However, due to noise, there are fluctuations around this mean value which are indicated by dashed arrows.}
  \label{fig-schematic}
\end{figure}

Another important mechanism commonly studied in the context of developing embryos is the the French-flag model \cite{WOLPERT19691, sharpe2019}. \redw{In order to build complex body structures, individual cells have to take decisions and adopt fates suitable for their positions. Yet cells do not have any direct way to measure their positions.} The key idea in the French-flag model is that during the early developmental stage of an embryo, some specific chemicals, called `morphogens' are deposited locally on one side of the embryo. \ps{These chemicals diffuse inside the embryo, thereby establishing a concentration gradient \cite{Gregordiffusion, Grieneisen2012}.} Unlike in Turing patterns, the spatial symmetry is broken due to the presence of graded concentration of morphogens. Wolpert proposed that cells could determine their positions from the local morphogen concentrations within these graded profiles and take up fates correlated with their positions \cite{WOLPERT2016597, WOLPERT19691}. Therefore, the graded-concentration profile of signalling morphogen provides `positional information' to the cells, see  Figure \ref{fig-schematic}. Cells, in turn, read out these positional cues in a threshold-dependent manner and give rise to spatial patterns with distinct boundaries. %This mechanism has been observed in the segmented anterior/posterior patterning in Drosophila embryo where a gradient in morphogen Bicoid leads to the expression levels of various genes being correlated with their physical positions \bluew{\cite{..}}. 

\ps{Although the idea of positional information was originally introduced in the context of developmental biology, its applications reach beyond this. For example, synthetic versions of the French-Flag model have been constructed and positional information can play a crucial role in the construction of self-assembling soft materials, where individual components could determine their position through the chemical concentration of a signalling molecule \cite{Zadorin2017, BACCOUCHE2014234,toda2020engineering,dupin2022synthetic}.}

Recently a mathematical framework has been developed to quantitatively study positional information \cite{Dubius2013, Gregor2007, Bialekbook2012,Gasper2014,Gasper2016, Gasper2021, MRao2023}. Essentially, one defines the positional information as the amount of information one gets about the position by measuring the concentration of signal molecules in this position. \ps{Based on the information theory, this amount of information is equal to}
%: During the initial stage of development, cells inside an embryo have similar structures, and if one does not measure the expression levels in a cell, it could lie anywhere in the embryo. In the language of probabilities, this means that the position $x$ of a cell is drawn from a prior probability distribution $P_x(x)$, which is often chosen to be uniform. Now if one observes a certain expression level in a cell (which say corresponds to a morphogen concentration $n$), then the position of a cell can be more accurately specified. However, due to the fact that the cells are in a noisy environment [see Figure \ref{fig-schematic}], the position of a cell after measurement is still drawn from a distribution $P(x|n)$ conditioned on the value $n$. Note that this conditional distribution is always narrower than the prior probability $P_x(x)$, and the degree of narrowness represents the positional information gained. Using concepts from information theory, it was demonstrated in \bluew{\cite{..}} that the positional information can be represented as
%\begin{align}
%\mathcal{I}_{(n , x)} \left( \{\rho\}  \right) & = \int dn~dx~ P(x,n) \log_2 \left[\frac{P(x,n)}{P_x(x) P_n(n)} \right], \label{avg-information-new}
%\end{align}
\begin{align}
\mathcal{I}_{(n , i)} \left( \{\rho\}  \right) & = \sum _{i}~\int dn~ P(i,n) \log_2 \left[\frac{P(i,n)}{P_i(i) P_n(n)} \right], \label{avg-information-new}
\end{align}
where $P(i,n)$ is the joint probability distribution observing a concentration $n$ at position $i$ and $P_i(i)$ and $P_n(n)$ are the associated marginals. \textcolor{black}{Throughout our paper, we have associated every lattice point with an elementary volume $v_c$, c.f., Fig.~\ref{avg-information-new}. Such a consideration may arise because, in biological experiments, the cells for which positional information is measured have a finite volume that is small compared to the overall size of the embryo.}

%Such a discretisation might arise because in biological experiments cells for which positional information is measured, have a finite volume which is small compared to that of the entire embryo.Typically in biological experiments, cells for which positional information is measured, have a finite volume might arise due to the finite measurement resolution in biological experiments, where positional information is generally measured for cells that have a finite volume.

\ps{Coming to our formula in Eq.~\eqref{avg-information-new}, the prior distribution $P_i(i)$ is chosen based on the the initial knowledge of the system while other probabilities are uniquely determined through the system's dynamics.} We also remark that if $n$ takes only discrete values, then the integration in Eq.~\eqref{avg-information-new} needs to be replaced appropriately by its summation. Also, we have written positional information as a function of model parameters $\{ \rho \}$ whose precise definition will be given later.

%\textcolor{purple}{In experiments involving early embryonic development in Drosophila, most cells have the same morphology specially in the central ($\sim 80 \%$) region of embryo \citep{Dubius2013,PRXLife.2.013016, Gasper2021}. Moreover, they are uniformly distributed \cite{PRXLife.2.013016}. If one does not perform any measurement on these cells, they could lie anywhere along the anterior/posterior axis. Without any prior knowledge, it is therefore natural to set $P_i(i)$ to be uniform \cite{MRao2023, Dubius2013,PRXLife.2.013016, BACCOUCHE2014234}. However, this is not an essential requirement for our framework.}

%\ps{In experiments involving early embryonic development in Drosophila, most cells have the same morphology specially in the central ($\sim 80 \%$) region of embryo \citep{Dubius2013,PRXLife.2.013016, Gasper2021}. Moreover, they are uniformly distributed \cite{PRXLife.2.013016}. If one does not perform any measurement on these cells, they could lie anywhere along the anterior/posterior axis. This has led to the experimental measurements of the positional information based on the uniform prior assumption  \citep{MRao2023, Dubius2013,PRXLife.2.013016, BACCOUCHE2014234}.  While this is a natural assumption in experiments, it is not a necessary requirement for our framework.However, this is not a necessary assumption for our framework and the results can be extended to the non-uniform case (also discussed later).} 

\ps{This framework has been successfully applied to biological systems. For example, experiments reveal that the four gap genes in the Drosophila embryo carry approximately $\sim 4.2$ bits of information. This amount of information enables cells to know their positions within an error of $\sim 1 \%$ of the total embryo length \cite{Dubius2013}. Since position determines a cell's fate, such precision is crucial for the robust and reproducible positioning of body structures like stripes of gene expression or cephalic furrow, while the system functions in a noisy environment. To understand how this developmental robustness is achieved, it is important to know how much information can be carried by the morphogens, but there are, to the best of our knowledge, no results on the fundamental physical limits of how much positional information can be stored inside a system.}
%Such level of precision is critical for the development of robust body structures, such as cephalic furrow whose positioning is reproducible with an accuracy of about $\sim 1 \%$ of the total embryo length \bluew{\cite{..}}. Furthermore, a multitude of morphogen concentration profiles, both linear and non-linear, have been observed in various organisms and tissues \bluew{\cite{..}}. 

%In order to set up positional information, morphogen molecules involve non-equilibrium transport process from one side of the embryo to the other. As such, we expect the positional information to strongly depend on to what extent the system deviates from equilibrium. Is there any general quantitative relation between them? Given the importance of positional information in pattern formation, answering this question is of paramount importance. 
The central goal of this paper is to fill this gap by determining the maximal amount of positional information stored in \bluerev{symmetric} boundary-driven \bluerev{discrete lattice} systems \bluerev{, i.e., systems that are in equilibrium in the absence boundary driving and that satisfy translational and left-right symmetry}, in terms of their distance from thermodynamic equilibrium \cite{Blythe_2007, BDerrida_1993, PhysRevLett.87.150601, Derrida2002, Derrida_2007, Levine2005, SCHUTZ20011}. In particular, we will look at systems that are in contact with two particle reservoirs with chemical potentials $\mu _L$ and $\mu _R$ [and $\Delta \mu = (\mu_L-\mu_ R$) being their difference]. \ps{We show that within linear response around equilibrium and under a local-equilibrium assumption,  positional information takes a universal form in terms of the chemical potentials and  densities at the left and right reservoirs
\begin{align}
\mathcal{I} _{\left( n, i\right) } \left( \bar{\rho}, \Delta \rho \right) \approx  \frac{v_c~\Delta \mu ~\Delta \rho }{ k_B T\ln2} ~\text{var}\left( \frac{i}{N} \right), \label{psbqpj8q}
\end{align}
where we have $\bar{\rho} =( \rho _L + \rho _R)/2$ and $\Delta \rho = (\rho _L - \rho _R)/2$, with $\rho _L$ and $\rho _R$ the densities at the left and right reservoir respectively, $k_B$ is the Boltzmann constant and $T$ is the absolute temperature at which the system is kept. \textcolor{black}{Furthermore, the volume $v_c$ is an elementary volume associated with every lattice point in the system},  and $\text{var} \left( f_i\right)$ indicates the variance of a function $f_i$ with respect to $P_i(i)$ and $N~(\gg 1)$ is the total number of discrete positions, c.f.~Fig.~\ref{avg-information-new}. Although $\bar{\rho}$ does not explicitly appear in Eq.~\eqref{psbqpj8q}, its influence is solely through $\Delta \mu$, which depends on both $\bar{\rho}$ and $\Delta \rho$.\\
\indent
In the far-from-equilibrium regime, we numerically find that the equality of Eq.~\eqref{psbqpj8q} is replaced by an inequality:
\begin{align}
  \mathcal{I} _{\left( n, i\right) } \left( \bar{\rho}, \Delta \rho \right) \leq \frac{v_c~\Delta \mu ~\Delta \rho }{ 4 ~ k_B T \ln2},  \label{aibdi172}
\end{align}
While this can be rigorously proven for any system close to equilibrium, we numerically find it to be valid even in the far-from-equilibrium condition for a broad class of systems \cite{SSMSingh}. This inequality thus provides a quantitative link between the positional information and the drive to maintain non-equilibrium and for a given driving, it tells us that there is a limit on how much positional information can be generated. \\
\indent
Although results~\eqref{psbqpj8q} and \eqref{aibdi172} hold for any prior, both experiments in biological systems and synthetic biological experiments generally use a uniform distribution of cells. Experimentally, one therefore generally considers a flat prior, $P_i(i)=1/N$ \cite{MRao2023, Dubius2013,PRXLife.2.013016, BACCOUCHE2014234, Gasper2021}. For this case, we demonstrate numerically a tighter upper bound to the positional information compared to \eqref{aibdi172}
\begin{align}
\mathcal{I} _{\left( n, i\right) } \left( \bar{\rho} , \Delta \rho \right)  \leq \frac{v_c~\Delta \mu~\Delta \rho  }{ 12 ~k_B T \ln 2}. \label{ineqq-eq-1}
\end{align}
where the equality is reached in the near-equilibrium condition, can be seen from Eq.~\eqref{psbqpj8q} by noting $\textrm{var}(i/N) \simeq 1/12$ for large $N$.}\\~\\
%\begin{align}
%\mathcal{I} _{\left( n, i\right) } \left( \bar{\rho} , \Delta \rho \right) \approx \frac{v_c~\Delta \rho ~\Delta \mu }{ 12 ~k_B T \ln 2 }. \label{psbqpj8q} 
%\end{align} 
%In the far-from equilibrium regime, we demonstrate numerically that this relation provides a tighter upper bound to the positional information compared to \eqref{aibdi172} for the uniform prior case

\noindent 
\textit{Positional information near equilibrium:} We begin with a one-dimensional lattice system consisting of $N$ sites represented by index $i$ with $1 \leq i \leq N$. 
%These sites can either accommodate an arbitrary number of particles (as observed in systems like ZRP or IRW) or can have at most a fixed number of particles (as seen in SSEP). 
In bulk where $1 < i < N$, a particle can jump to either of its neighbouring sites with an arbitrary rate (the rate can also be a function of the number of particles in the sites). On the other hand, at the two end sites ($i=1$ and $i=N$), the system is in contact with two particle reservoirs characterised by the average densities $\rho _L$ and $\rho _R$. Without any loss of generality, we will take $\rho _L \geq \rho _R$ throughout this paper.
\\
\indent
Due to the coupling with the reservoirs, our lattice system eventually reaches a non-equilibrium steady-state. Our starting point is to note that even though the system as a whole is driven out-of-equilibrium, local regions can still be described, to first order in gradients, by an equilibrium measure with parameters that vary slowly across the system \cite{Groot1984, PhysRevLett.112.100602}. This \textit{local thermodynamic equilibrium} then allows us to identify thermodynamic quantities such as chemical potential locally even in out-of-equilibrium set-ups. With this idea in mind, we now consider a small volume $v_c$ around the lattice site $i$ and write the probability distribution to observe a number $n$ inside this volume in the steady-state as
%With this idea in mind, one can introduce a scaled continuous variable $x =i/N$ and take the limit $N \rightarrow \infty$, so that $x \in [0,1]$ and write the distribution to observe a local density $n(x)$ at position $x$ in the steady-state as \cite{Derrida_2007}
\begin{align}
P(n|i) \sim  \exp \left[ - \frac{v_c}{k_B T}\left( \mathcal{G}_{\mu _i}\left( n/v_c\right) - \mathcal{G}_{\mu _i}\left(  \rho _i \right) \right) \right],  \label{lett-general-bd-eq-3}
\end{align}
where $\mathcal{G}_{\mu_i}\left( g\right)$ is given in terms of the Helmholtz free energy $a(n)$ per unit volume as
\begin{align}
\mathcal{G}_{\mu _i}(g) = a(g) - \mu _i g,~~\text{and } \mu _i = \frac{\partial a(g)}{\partial g} \Big|_{g =  \rho _i  }. \label{lett-general-bd-eq-2}
\end{align}
Here $\mu _i $ stands for the local chemical potential which is given in terms of the local average density $\rho _i = \langle n _i/v_c \rangle $ such that $\rho_1 = \rho _L$ and $\rho_N = \rho _R$. \ps{In this description, the microscopic details of the system such as inter-particle interaction, dynamics etc, are captured by the form of free energy $a(g)$ and the average density $\rho _i$. Their forms will depend on the specific model \cite{SSMSingh}. \textcolor{black}{Furthermore, although we have examined discrete lattice systems in this paper, their continuum version can be derived by coarse-graining the microscopic model \cite{Derrida2002, Derrida_2007}. In this context, $v_c$ represents the coarse-graining length scale.}}

In general, the form of the density $\rho _i$ can be both linear or non-linear in $i$. However, close to equilibrium, one generally expects it to take a linear form $\rho_i \simeq \bar{\rho} +\left(1-\frac{2i}{N} \right) \Delta \rho$ according to the diffusion equation. This is also in agreement with experimental observations for certain morphogens \cite{Thompson2023}. Plugging this in Eq.~\eqref{lett-general-bd-eq-3} then gives us the conditional distribution $P(n|i)$. Recall that we also need the joint probability distribution $P(n,i)$ in Eq.~\eqref{avg-information-new}. Using Bayes' theorem, we can write $P(i,n) = P(n|i) P_i(i)$. \ps{We now have all quantities needed in Eq.~\eqref{avg-information-new} and we show in \cite{SSMSingh} that the positional information is given by
\begin{align}
\mathcal{I} _{\left( n, i\right) } \left( \bar{\rho}, \Delta \rho \right)    \approx &~ \frac{2\Delta \rho ^2}{\ln2~\sigma _2(\bar{\rho})} ~\text{var}\left( \frac{i}{N}\right), \label{mabska}
\end{align}}
where \ps{$\sigma _2(\bar{\rho} ) = \langle (n-\bar{\rho})^2\rangle _{\text{eq}}$} is the variance of the density at equilibrium $(\Delta \rho=0)$.  As expected, $\mathcal{I} _{\left( n, i\right) } \left( \bar{\rho}, \Delta \rho \right) $ goes to zero at equilibrium since the density remains constant across the system and therefore the chemical concentration does not correlate with position.
%\textcolor{purple}{For this, we need the knowledge of $P_i(i)$. Here, we will present our derivation for the case of a flat prior $P_i(i) = 1/N$, while the extension to non-flat priors is provided in the supplemental materials \cite{SSMSingh}.}
\begin{figure}[t]
\centering
  \includegraphics[scale=0.35]{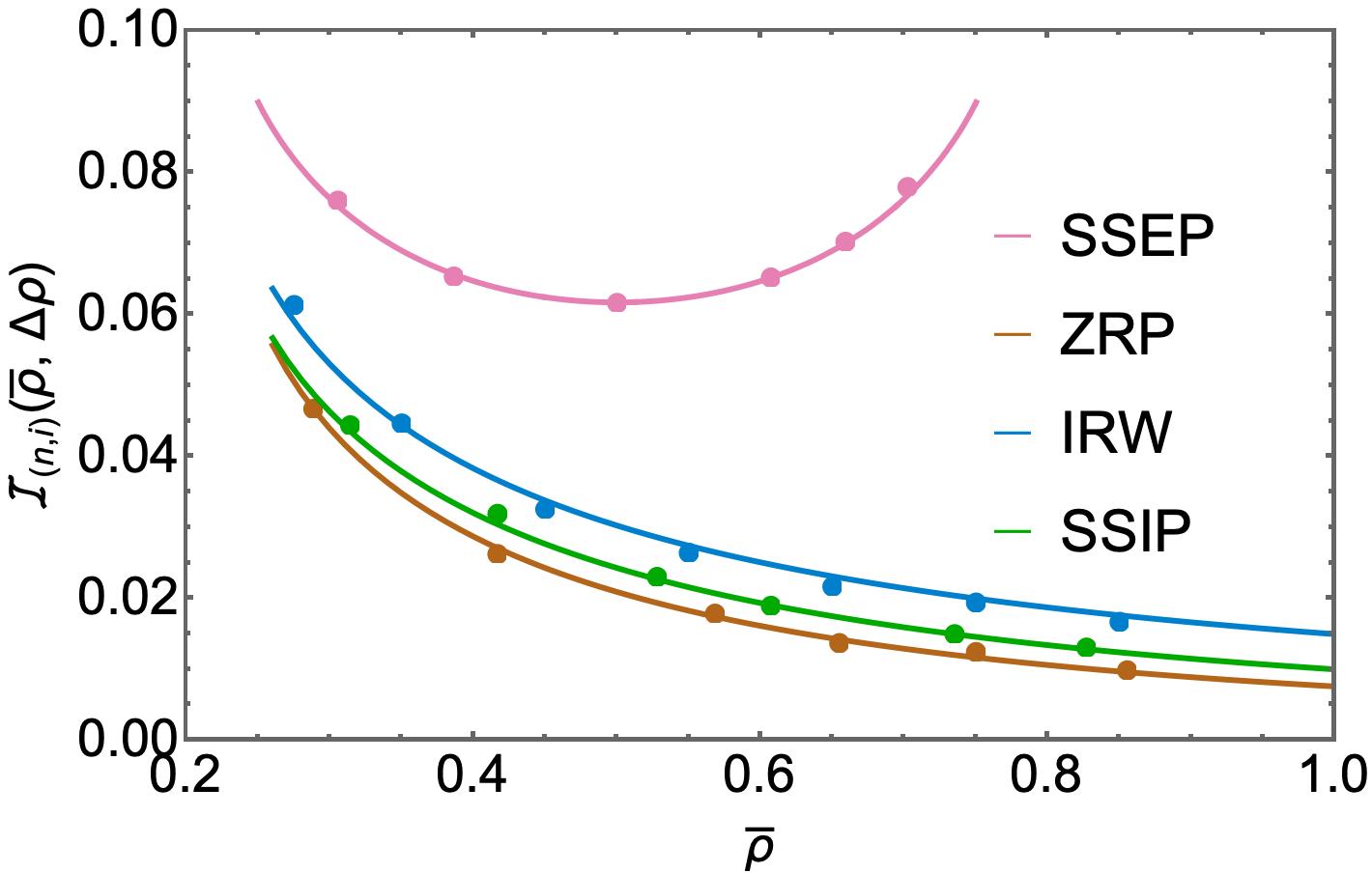}
  \caption{Plot of the positional information as a function of $\bar{\rho}$ for four different boundary-driven processes and its comparison with the numerical simulations.  In each case, solid lines illustrate the analytical expression, c.f., summary table I in \cite{SSMSingh}, while symbols denote simulation data. We have fixed the bias to $\Delta \rho = 0.25$ \ps{and chosen a flat prior.}}
  \label{fig-1}
\end{figure}
We now establish the connection of positional information with the chemical potential difference $\Delta \mu = (\mu _L-\mu_R)$. Noting Eq.~\eqref{lett-general-bd-eq-2}, this can be expressed in terms of the derivative of the free energy. At the leading order in $\Delta \rho$, we find $\Delta \mu \sim a_2(\bar{\rho})~\Delta \rho$, where $a_2(\bar{\rho})$ stands for the second derivative of $a(\bar{\rho})$. We next use the fluctuation-response relation to write $a_2(\bar{\rho}) =k_B T/ \sigma _2(\bar{\rho}) v_c $; see \cite{SSMSingh} for a proof. This leads us to the expression of chemical potential difference as
\begin{align}
\frac{\Delta \mu }{k_B T} \approx \frac{2 \Delta \rho}{v_c\sigma _2 (\bar{\rho})}. \label{mabskagaiq}
\end{align}
\ps{Combining this with Eq.~\eqref{mabska}, we arrive at the universal form of the positional information quoted in Eq.~\eqref{psbqpj8q}. Moreover, we prove in \cite{SSMSingh} that the variance is bounded as $\textrm{var}(i/N) \leq 1/4$ and applying this in Eq.~\eqref{psbqpj8q}, we obtain the general upper bound \eqref{aibdi172}. These are the two main results of our letter. They are valid in the near-equilibrium regime for any $1$-dimensional boundary-driven system with arbitrary prior probability. For the biologically relevant case, $P_i(i)=1/N$, we numerically observe that the linear response result in Eq.~\eqref{psbqpj8q} provides a further tight bound \eqref{ineqq-eq-1} to the positional information in the far-from-equilibrium regime.} \\
%numerically find a further refined bound in \eqref{ineqq-eq-1}. saturate the universal upper bound on the positional information as given in \eqref{ineqq-eq-1}, by noting that \textcolor{purple}{$\textrm{var}(i/N)\simeq 1/12$ for large $N$}.
%In fact, it turns out that one can utilize this framework beyond linear response regime and for non-linear density profiles also. For this,  we need the expansion  $\rho (x)  = \bar{\rho} + \sum _{k=1}^{\infty} \mathcal{L}_k(x, \bar{\rho})~\Delta \rho ^k$, where the $\mathcal{L}_k(x, \bar{\rho})$ functions vary depending on the model. With this, the positional information is obtained as a series in $\Delta \rho$ with prefactor expressed in terms of $\sigma _k(\bar{\rho})$ and $\mathcal{L}_k(x, \bar{\rho})$ functions [see \bluew{\cite{..}} for details].
%\begin{table}[H]
%\centering
\begin{figure}[t]
\centering
  \includegraphics[scale=0.16]{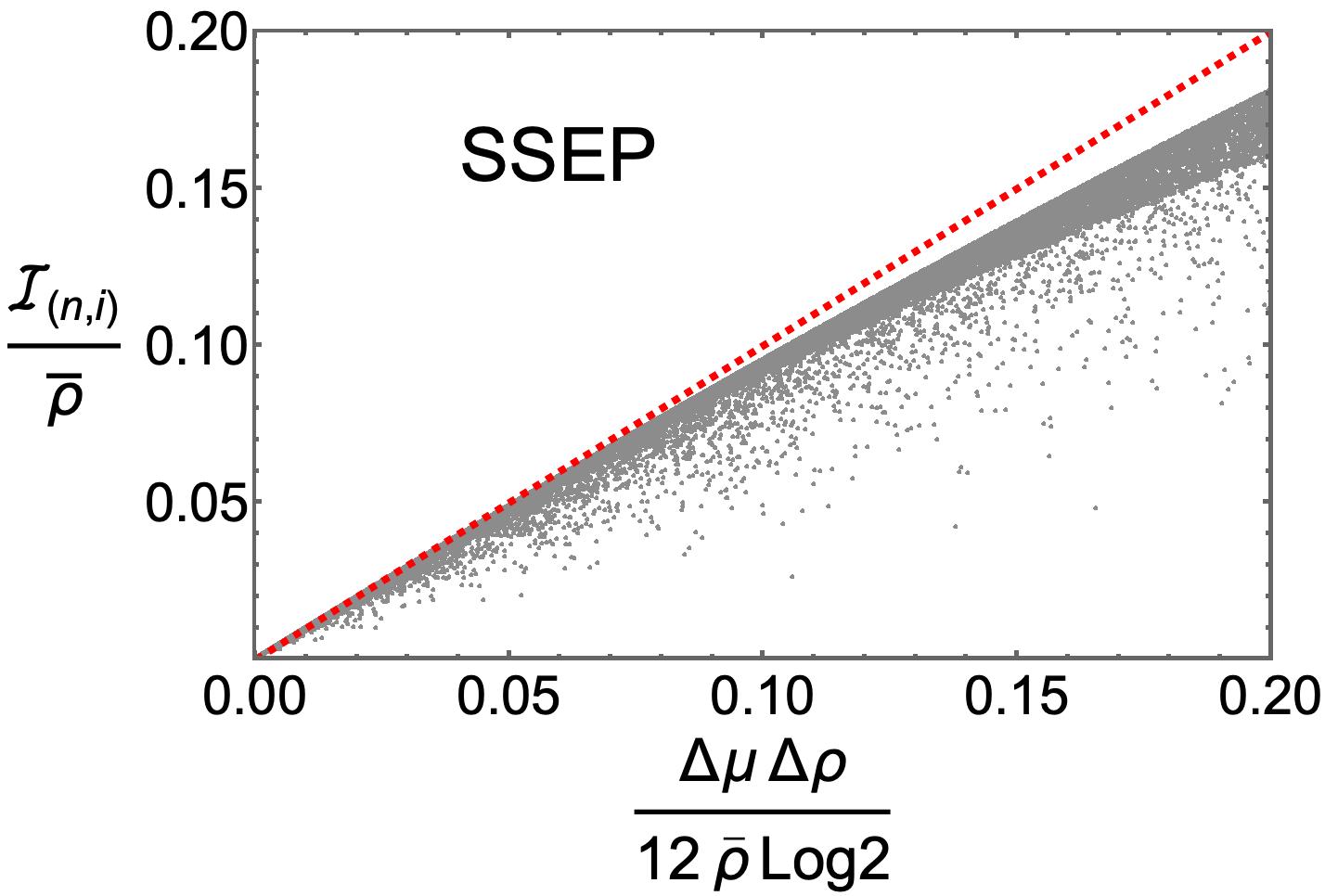}
   \includegraphics[scale=0.16]{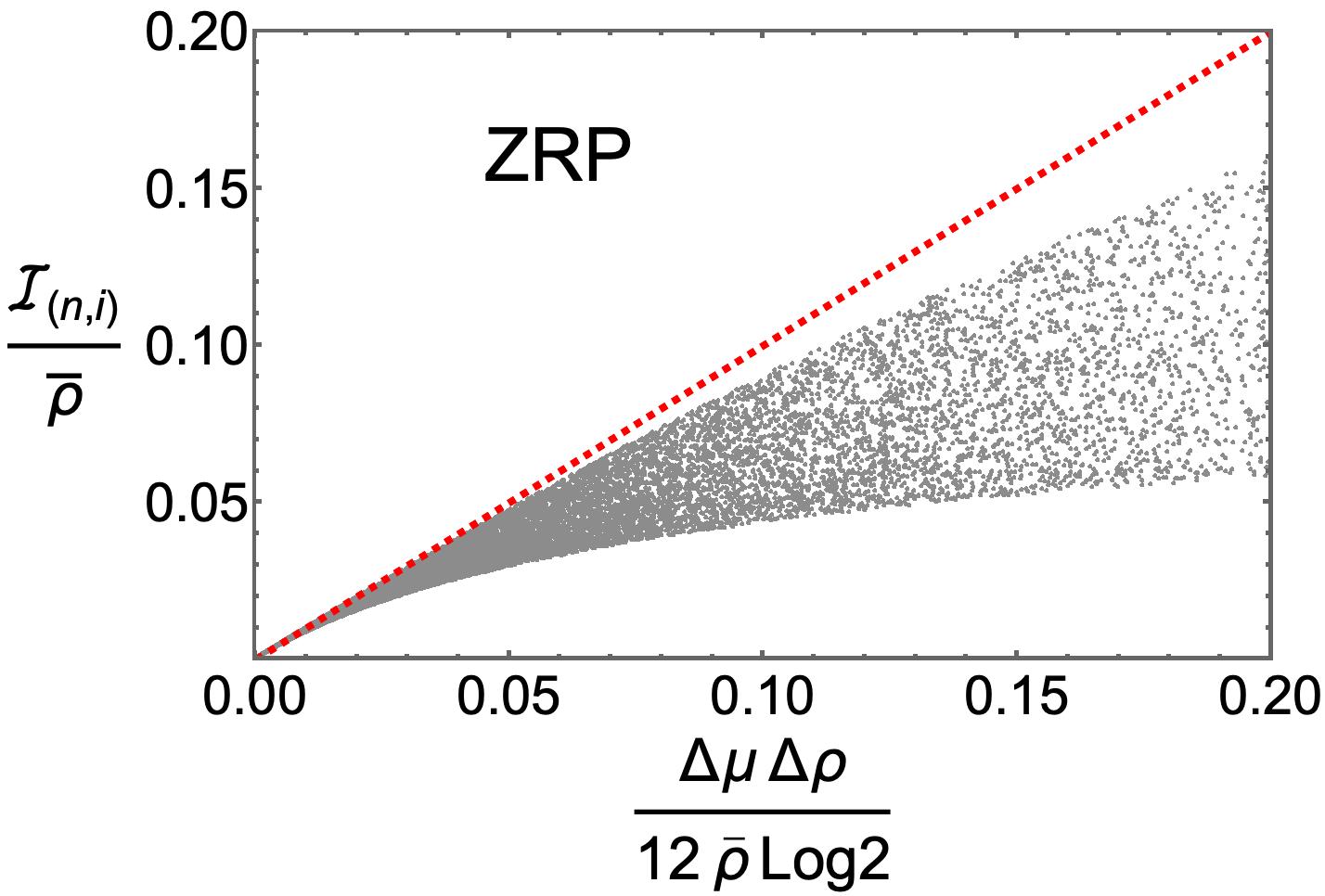}
  \includegraphics[scale=0.16]{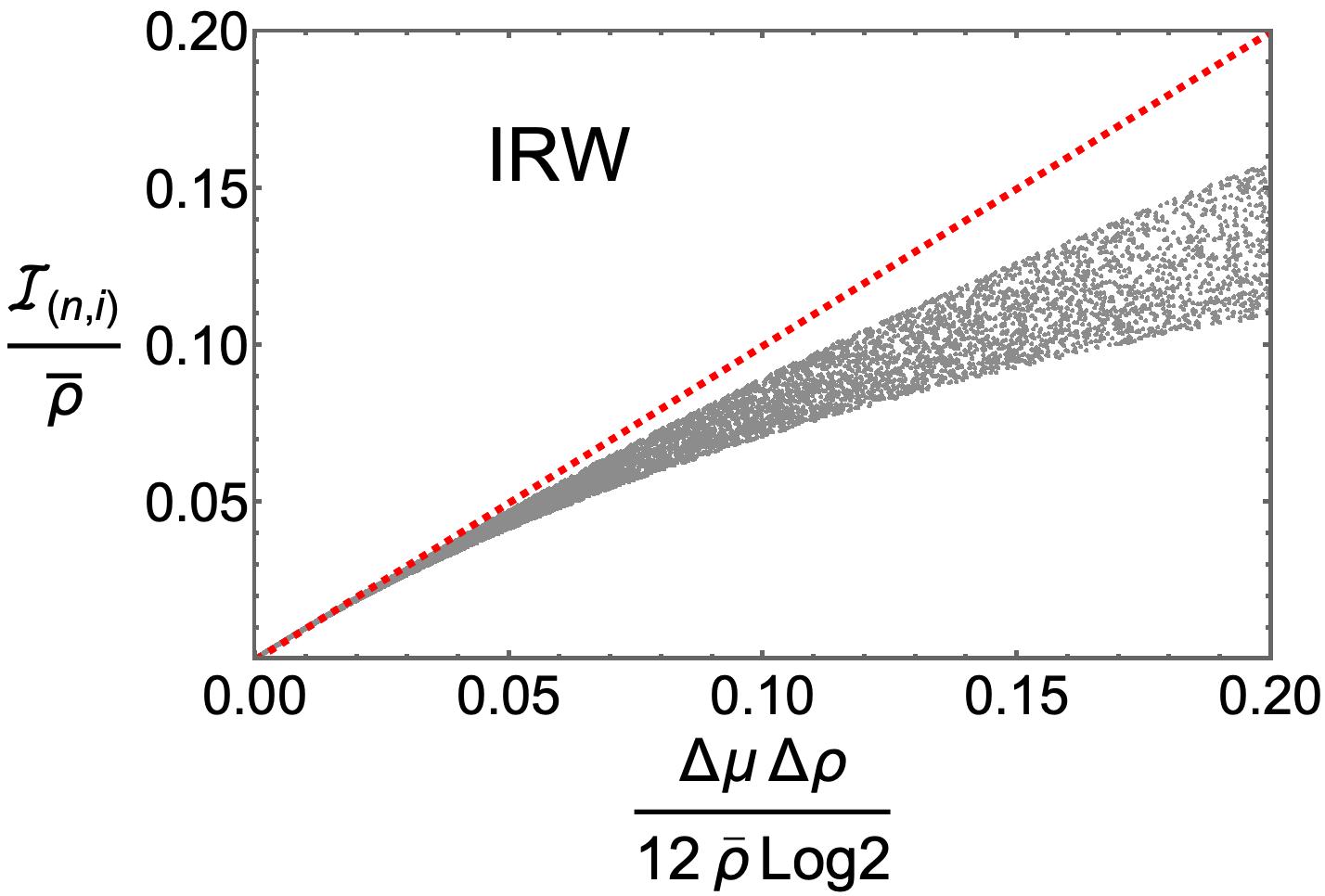}
  \includegraphics[scale=0.16]{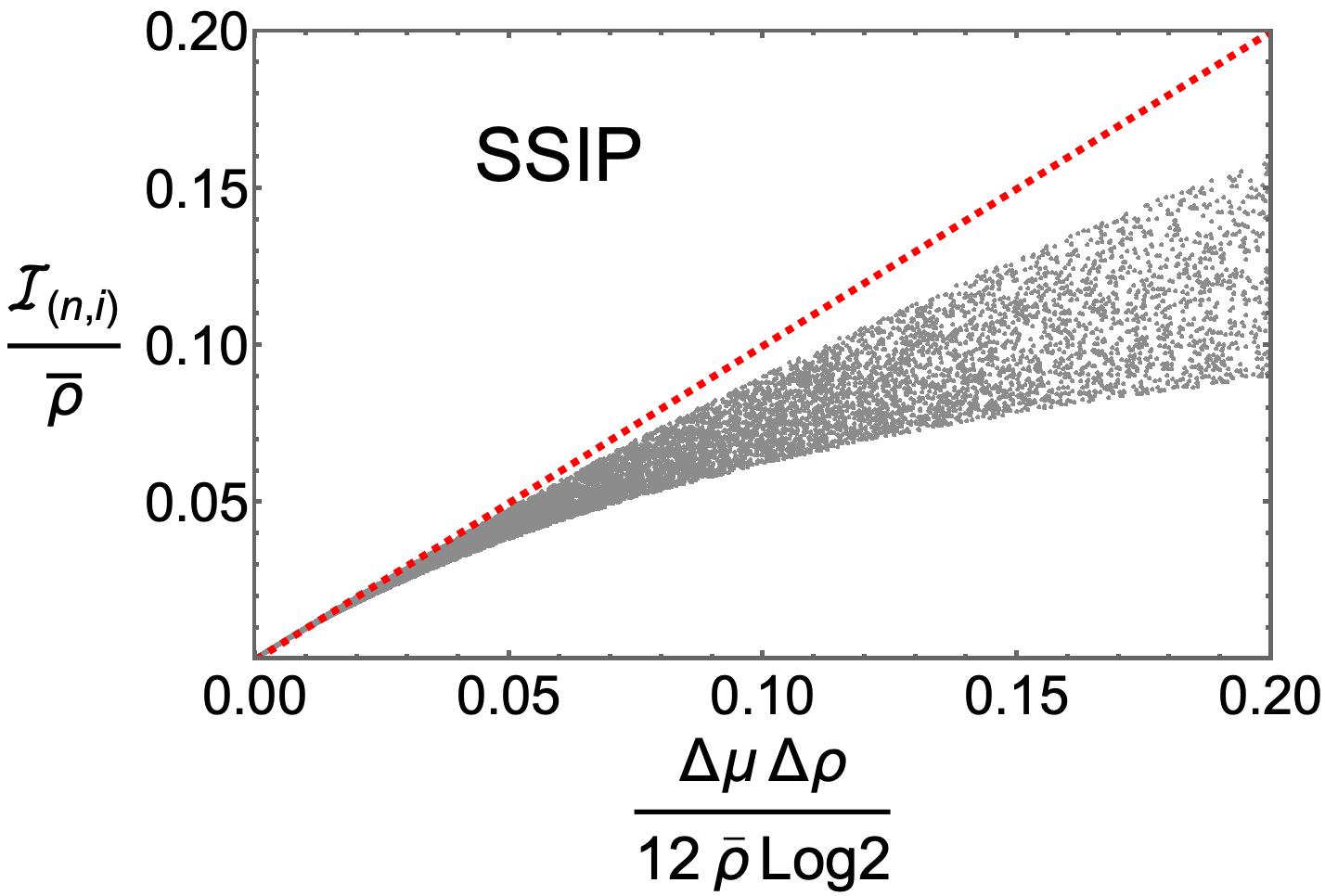}
  \caption{\ps{Upper bound \eqref{ineqq-eq-1} on the positional information in terms of the differences $\Delta \mu$ and $\Delta \rho$ for four different boundary-driven processes with a flat prior. For all cases, the gray symbols are the analytical results for different random values of $\bar{\rho}$ and $\Delta \rho$ generated using the procedure explained in \cite{SSMSingh}. For better visibility, we have rescaled both sides of \eqref{ineqq-eq-1}  by $\bar{\rho} $ and plotted $\mathcal{I} _{\left( n, i\right) } \left( \bar{\rho}, \Delta \rho \right) / \bar{\rho} $ vs $\Delta \mu~\Delta \rho / 12 \bar{\rho} \ln 2$. As discussed in \cite{SSMSingh}, this rescaling ensures that the near-equilibrium relation \eqref{psbqpj8q} is more clearly visible.
The red line corresponds to the upper bound in \eqref{ineqq-eq-1}. Clearly, the bound is saturated in the near-equilibrium limit. We have set $v_c=1$ and $k_B T=1$.}}
\label{ratio-fig}
\end{figure}
\noindent
\textit{Far-from-equilibrium regime:} Up to now, our analysis has focused on the near-equilibrium situation where local thermodynamic equilibrium enabled us \ps{to derive a general bound. An immediate question that now follows is - what happens to this bound in the far-from-equilibrium situations? Although we are unable to give an analytical proof, numerical results on a broad range of models suggest it remains valid even when the system is driven arbitrarily far from equilibrium. We will now show this for four different models, namely symmetric simple exclusion processes (SSEP), zero-range processes (ZRP), independent random walkers (IRW), and simple symmetric inclusion processes (SSIP). In the rest of our analysis, we will focus on the flat case $P_i(i) =1/N$ and instead test the refined bound \eqref{ineqq-eq-1} for these solvable examples. We verify \eqref{aibdi172} in the supplementary materials \cite{SSMSingh}. Furthermore, we will set $v_c$ to the volume associated with a single lattice site and write $k_B T=1$ and $v_c=1$ in the remainder of our letter for notational simplicity.}\\
\indent
\textit{Example I: SSEP-} Consider the SSEP model where every lattice site can either be vacant $(n_i =0)$ or be occupied by a single particle $(n_i =1)$. Any particle in the bulk can jump to one of its neighbouring sites with some rate $p$, as long as the target site is vacant. However, dynamics at the end sites are modified due to the presence of particle reservoirs, see \cite{SSMSingh} for details. Since the occupation number $n_i$ for this model is a binary variable, one can write the conditional probability $P(n|i) = \rho _i \delta _{n,1}+ \big[ 1-\rho _i \big] ~\delta _{n,0}$ with density $\rho _i = \bar{\rho} +\left(1- \frac{2i}{N} \right)\Delta \rho$ \cite{BDerrida_1993, PhysRevLett.87.150601, Derrida2002, Derrida_2007}. If we now use this in Eq.~\eqref{avg-information-new}, the positional information can be calculated. We have presented this expression in summary table I in \cite{SSMSingh}, and validated it against numerical simulations in Figure~\ref{fig-1}. It is also consistent with our general result in Eq.~\eqref{psbqpj8q} in the limit $\Delta \rho \to 0$, as one can show that $\sigma _2(\bar{\rho}) = \bar{\rho}(1-\bar{\rho})$. Our objective now is to compare this expression with the chemical potential difference $\Delta \mu $ quoted in summary table I in \cite{SSMSingh}. In Figure~\ref{ratio-fig}, we have plotted both $\mathcal{I} _{\left( n, i\right) } \left( \bar{\rho} , \Delta \rho \right)$ and $\Delta \mu \Delta \rho / 12 \ln 2 $ for several values of $\bar{\rho}$ and $\Delta \rho $. The red dashed line represents the bound in \eqref{ineqq-eq-1}. Clearly all positional information values lie below this line. In fact, for SSEP, we have rigorously proved in \cite{SSMSingh} that the bound is valid across all parameter values. Hence, our upper bound \eqref{ineqq-eq-1} holds for the SSEP arbitrarily far from the equilibrium.\\
\indent
\textit{Example II: ZRP-} As a second example, we look at the boundary-driven zero-range process \cite{Levine2005}. \ps{Unlike in SSEP, the lattice sites are now capable of accommodating an arbitrary number of particles.} In bulk, a particle can jump to any of its neighbouring \ps{sites} with a rate of $p u_{n_i}$, where $u_{n_i}$ is a non-negative function of $n_i$. At the boundaries, the dynamics are modified allowing for the addition or removal of particles. 
%For some choices of $u_{n_i}$, it is known that the system develops a condensate \bluew{\cite{..}}. Here, we avoid this complexity and focus on regimes that do not have condensation.
For this model, the steady-state density profile is non-linear. Moreover, for the case of constant rate $u_{n}=1$, the positional information can be calculated as shown in \cite{SSMSingh}. In Figure~\ref{fig-1}, we have compared this expression with numerical simulations and found a good agreement between them. Next we employ this result in conjunction with $\Delta \mu$ to test the upper bound \eqref{ineqq-eq-1}.
%Contrary to SSEP, we observe that the positional information for ZRP decreases monotonically with $\rho _L$ and vanishes in the limit $\rho _L \to \infty $ The monotonicity form for SSEP is a consequence of the exclusion dynamics, and is absent in models with non-exclusion particle dynamics (cf. Figure~\ref{fig-1})\\
%We now compare Eq.~\eqref{ZRP-paper-eq-1} with our bound \eqref{ineqq-eq-1}. 
As depicted in Figure~\ref{ratio-fig}, positional information values are situated below the red line even for this model. This observation holds true across all parameter regimes. The bound is saturated in the near-equilibrium regime where both $\Delta \mu$ and $\Delta \rho$ approach zero as expected. Furthermore, in this example also, $\mathcal{I} _{\left( n, i\right) }\left( \bar{\rho}, \Delta \rho \right)$ is bounded by \eqref{ineqq-eq-1} even in the far-from equilibrium conditions.
\\
\indent
\textit{Example III: IRW-} Another choice of rate $u_n=n$ corresponds to the independent, unbiased random walkers \cite{Levine2005} and turns out to be analytically solvable. Once again, we use the positional information and chemical potential difference $\Delta \mu$ from \cite{SSMSingh} to check our bound for this model. Plotting $\mathcal{I} _{\left( n, i\right) } \left( \bar{\rho} , \Delta \rho \right) $ and $  (\Delta \mu \Delta \rho )/( 12 \ln 2)$ in Figure~\ref{ratio-fig}, we have demonstrated that the upper bound is satisfied for this model. This model presents a third solvable example where the upper bound is valid in all parameter regimes.\\
\indent
\textit{Example IV: SSIP-} As a final example, we investigate the simple symmetric inclusion process \cite{PhysRevE.90.052143, 10.3150/21-BEJ1390}. For this model also, a lattice site can accommodate an arbitrary number of particles. However unlike the previous examples, the jump rate from $i \to j$ [with $j=(i \pm 1)$] depends on the both occupation numbers $n_i$ and $n_j$. We choose the form of this rate as $p n_i(n_j+m)$ where $m~(>0)$ is a parameter in the model. Furthermore, the system is coupled with two reservoirs at the boundaries which drive it to a non-equilibrium steady-state. In this steady-state, we calculate the positional information and $\Delta \mu$ as shown in \cite{SSMSingh}. These results are  plotted in the bottom panel of Figure~\ref{ratio-fig}. From this, it is evident that the upper bound \eqref{ineqq-eq-1} is satisfied by this model. This further reinforces the conjecture that our bound  holds for general boundary-driven system. \redw{Therefore, in all the examples investigated in this letter, we observe that there exists a quantitative constraint on the amount of positional information given in terms of the chemical potential and density gradients driving the system.}\\
\indent
\textit{Conclusions and outlook:} In conclusion, we established an important link between biophysics and non-equilibrium statistical mechanics, by studying the limits of positional information. \ps{More specifically, we derived a universal expression for the positional information $\mathcal{I} _{\left( n, i\right) } \left( \bar{\rho}, \Delta \rho \right)$ near equilibrium, involving only the chemical potential difference and the difference in average densities driving the system. Furthermore, our analysis on several solvable models suggests that this expression serves as an upper bound on positional information in the far-from-equilibrium regime. In particular, for the case of flat prior, we find  $\mathcal{I} _{\left( n, i\right) } \left( \bar{\rho} , \Delta \rho \right)  \leq v_c \Delta \rho  \Delta \mu/ 12~k_B T \ln 2 $.} Our study indicates that there is a limit on how much positional information the morphogen particles can provide, depending on how far the system is from equilibrium, a conclusion that has profound implications for the construction of future synthetic biological systems. \ps{Given that in many experiments, the morphogen profile is measured along one dimension such as the anterior-posterior axis of the embryo \cite{Dubius2013,PRXLife.2.013016, Grieneisen2012} or the longitudinal axis in microchannels \cite{Zadorin2017, dupin2022synthetic},  we have focused on one-dimensional boundary-driven models and used the principle of local equilibrium to express the conditional probability in Eq.~\eqref{lett-general-bd-eq-3}. However, it is possible to extend this principle even for some higher dimensional systems and our results will also be valid for them \cite{SSMSingh}.}

Our work paves the way for several future directions. While we examined specific models to demonstrate the bound, it remains an open problem to prove it generically. Extending this study for other models with bulk drive is an important future direction. This is especially relevant because, in some experiments, morphogen molecules experience degradation effects inside the embryo \cite{Gregordiffusion, Grieneisen2012}. \bluerev{Furthermore, biological systems often operate in an active non-equilibrium environment, leading to interesting phenomena such as motility-induced phase-separation \cite{MIPS}.} Under these circumstances, the local equilibrium assumption is no longer valid, and one needs to develop new theoretical methodologies to tackle the problem \bluew{\cite{Eyink1996}}.

%\ps{Throughout this letter, we have focused on one-dimensional boundary-driven models for which local equilibrium is satisfied, allowing us to express the conditional probability in Eq.~\eqref{lett-general-bd-eq-3}. In many experiments, the morphogen profile is measured along one dimension: the anterior-posterior axis of the embryo \cite{Dubius2013,PRXLife.2.013016, Grieneisen2012} and the longitudinal axis in microchannels \cite{Zadorin2017, dupin2022synthetic}. During the early developmental stages of Drosophila, the positional cues along different major axes of the embryo are mostly independent \cite{one-dimen-1, one-dimen-2, PRXLife.2.013016}. Hence, it is natural to study positional information in one dimension. In supplementary information \cite{SSMSingh}, we show that our result also works in two dimensions when reflecting boundary conditions are used in the second direction.}\\
\indent
%\ps{Although, it is natural to choose a uniform prior from the experimental perspective, it is not a necessary assumption for our letter. In fact, close to equilibrium, it is possible to show that the positional information is universally bounded as %\begin{align}
%  \mathcal{I} _{\left( n, i\right) } \left( \bar{\rho}, \Delta \rho \right) \leq \frac{\Delta \mu ~v_c~\Delta \rho }{ 4 k_B T \ln2}, 
%\end{align}
%for any choice of prior, see \cite{SSMSingh}. Moreover, we have tested its validity for all four models, even when the system is significantly away from equilibrium.} 

We thank Jonas Berx for carefully reading the manuscript. The authors also acknowledge the support from the European Union’s Horizon 2020 research and innovation program under the Marie Sklodowska-Curie grant agreement No. 847523 `INTERACTIONS’ and grant agreement No.~101064626 `TSBC' and from the Novo Nordisk Foundation (grant No. NNF18SA0035142 and NNF21OC0071284).

%\lipsum[1-5]

%\nocite{Bib_new}
\bibliography{Bib_new}

\end{document}

%\title{Thermodynamic limit to positional information}
\title{Limits to positional information in boundary-driven systems}

\author{Prashant Singh and Karel Proesmans}
\email{prashant.singh@nbi.ku.dk}
%\address{International Centre for Theoretical Sciences, Tata Institute of Fundamental ~Research, Bengaluru 560089, India $^1$}
\address{Niels Bohr International Academy, Niels Bohr Institute,
University of Copenhagen, Blegdamsvej 17, 2100 Copenhagen, Denmark}
%\ead{karel.proesmans@nbi.ku.dk}
%\ead{prashant.singh@icts.res.in, saikat.santra@icts.res.in}
%\ead{anupam.kundu@icts.res.in}
\vspace{10pt}

\clearpage
\onecolumngrid
\setcounter{equation}{0}
\renewcommand{\theequation}{S\arabic{equation}}
\setcounter{figure}{0}
\renewcommand{\thefigure}{S\arabic{figure}}
\section*{Supplementary Material: Limits to positional information in boundary-driven systems}
\begin{center}
Prashant Singh and Karel Proesmans\\
\textit{Niels Bohr International Academy, Niels Bohr Institute,}\\
\textit{University of Copenhagen, Blegdamsvej 17, 2100 Copenhagen, Denmark}
\end{center}

In this supplementary note, we will present an extensive derivation of the results which were quoted in the main text of our letter. To begin with, it is useful to recall the mathematical framework of positional information developed in \cite{Dubius2013}

\section{Mathematical framework for positional information}
Wolpert's idea is that although cells do not have any direct way to measure their positions, they can still acquire positional information by reading out the local concentration of the signalling morphogen molecules \cite{WOLPERT19691}. \pss{During the initial stage of development, cells inside an embryo might be distributed in a certain way, and depending on the initial knowledge about their positions, one will have some form of prior probability $P_i(i)$.} Now if one observes a certain certain morphogen concentration $n$, then the position of a cell can be more accurately specified. However, due to the fact that the cells are in a noisy environment, the position of a cell after measurement is still drawn from a probability distribution $P(i|n)$ but conditioned on the value $n$. Observe that this conditional distribution is always narrower than the prior $P_i(i)$, and the degree of this narrowness represents the amount of positional information gained by the cells. For instance, if the distribution $P(i|n)$ has same flatness as $P_i(i)$, then not much information is gained about the position of the cell. On the other hand, if $P(i|n)$ is highly peaked at some value of $i$, then the position of a cell is determined with more precision. 

The two distributions $P_i(i)$ and $P(i|n)$ are the main ingredients in this framework and the information gained by measuring the concentration $n$ is given by
\begin{align}
\mathbb{I}_{n \to i}  = S[P_i(i)] -  S[P(i|n )], \label{appen-pos-inf-eq-newra}  
\end{align}
where $S[P_i(i)]$ and $S[P(i|n )] $ are the entropies associated with distributions $P_i(i)$ and $P(i|n )$
\begin{align}
S[P_i(i)] & = - \sum_{i=1}^{N} ~P_i(i) \log _2 \left[P_i(i) \right], \\
S[P(i|n)] &= -\sum_{i=1}^{N}~P(i|n) \log _2 \left[P(i|n) \right].
\end{align}
The fact that $P(i|n)$ is narrower than $P_i(i)$ implies that $S[P(i|n)]$ is smaller than $S[P_i(i)]$. Hence $\mathbb{I}_{n \to i}$ in Eq.~\eqref{appen-pos-inf-eq-newra} can take only non-negative values. Moreover, we will assume that $n$ is a continuous variable. When it takes discrete values, the integration over $n$ need to be replaced by its summation (same applies to $i$).

Now if we randomly choose a cell, then the morphogen concentration will be distributed as $P_n(n)$ and taking the average of Eq.~\eqref{appen-pos-inf-eq-newra} with this distribution gives us
\begin{align}
\mathcal{I}_{(n , i)} \left( \{\rho\}  \right) &= \int dn ~P_n(n) \Big[S[P_i(i)] - S[P(i|n )] \Big],   \label{appen-avg-information-new}  \\
& = \sum _{i=1}^{N}\int dn~ P(i,n) \log_2 \left[\frac{P(i,n)}{P_i(i) P_n(n)} \right]. \label{appen-avg-information-newnvt} 
\end{align}
Here $P(i,n) = P(i|n) P_n(n) = P(n|i) P_i(i)$ stands for the joint distribution of $n$ and $i$. Moreover,  we have written positional information as a function of model parameters $\{ \rho \}$ whose precise definition will be given later. Interestingly, the expression in Eq.~\eqref{appen-avg-information-newnvt} emphasizes that the average information is mutual, \emph{i.e.} the information gained about the position of a cell by measuring the morphogen concentration is on average same as the information gained about the morphogen concentration by measuring the position of a cell. We therefore have
\begin{align}
\mathcal{I}_{(n , i)} \left( \{\rho\}  \right) = \sum _{i=1}^{N} ~P_i(i)~\Big[ S[P_n(n)]-S[P(n|i)] \Big], \label{appen-avg-information}
\end{align}
with entropies measured in bits as
\begin{align}
S[P_n(n)] & = - \int  dn ~P_n(n) \log _2 \left[P_n(n) \right], \\
S[P(n|i)] &= -\int dn~P(n|i) \log _2 \left[P(n|i) \right].
\end{align}
In some of our calculations below, Eq.~\eqref{appen-avg-information} turns out to be more useful to calculate the positional information.

\section{A perturbative approach for positional information in boundary-driven systems}
\label{sppen-sec-pert}
In this section, we develop a perturbative approach to obtain the positional information for general boundary driven systems. Let us consider a one-dimensional lattice system consisting of $N~(\gg 1)$ sites represented by the index $i$ with $1 \leq i \leq N$. These sites can either accommodate an arbitrary non-negative number of particles (as observed in systems like ZRP or IRW) or can have at most a fixed number of particles (as seen in SSEP). In bulk where $1 < i < N$, a particle can jump to either of its neighbouring sites with an arbitrary rate (the rate can also depend on the occupation numbers of these sites). On the other hand, at the two end sites ($i=1$ and $i=N$), the system is in contact with two particle reservoirs characterised by the average densities $\rho _L$ and $\rho _R$ and chemical potentials $\mu _L$ and $\mu _R$. \greenw{However, both reservoirs have the same temperature $T$.}
Without any loss of generality, we will take $\rho _L \geq \rho _R$ and denote the Boltzmann constant by $k_B$. Due to the coupling with the reservoirs, we assume that the system eventually reaches a non-equilibrium steady-state.
 
Our approach relies on the fact that even though there are significant departures from equilibrium in the system as a whole, local regions can still be described, to first order in gradients, by an equilibrium measure with parameters that vary slowly across the system This \textit{local thermodynamic equilibrium} allows one to identify thermodynamic quantities such as chemical potential locally even in out-of-equilibrium set-ups. With this idea in mind, we now consider a small volume $v_c$ around the lattice site $i$ and write the probability distribution to observe a number $n$ inside this volume as
%With this idea in mind, we now define a scaled continuous variable $x = i /N$ $(N \gg 1)$ with $x \in [0,1]$ and write the probability distribution to observe a local density $n$ at position $x$ as
\begin{align}
P(n|i) \sim  \exp \left[ - \frac{v_c}{k_B T} \Big( \mathcal{G}_{\mu_i}\left( n/v_c\right) - \mathcal{G}_{\mu _i}\left(  \rho_i  \right) \Big) \right], \label{supp-general-bd-eq-3}
\end{align}
where $\mathcal{G}_{\mu_i}\left( g\right)$ is given in terms of the Helmholtz free energy $a(g)$ per unit volume as
\begin{align}
\mathcal{G}_{\mu _i}(g) = a(g) - \mu_i g~~\text{and } \mu _i = \frac{\partial a(g)}{\partial g} \Big|_{g =  \rho _i  }. \label{supp-general-bd-eq-2}
\end{align}
Here $\mu _i $ is the local chemical potential which is given in terms of the local average density $\rho_i = \langle n_i/v_c \rangle $ such that $\rho_1 = \rho _L$ and $\rho_N = \rho _R$. \ps{In this description, the microscopic details of the system such as inter-particle interaction, dynamics etc, are captured by the form of $a(g)$ and the average density $\rho _i$. For models studied in the letter, the derivation of $a(g)$ is given in section~\ref{sec-appen-FE}}.

Although the precise form of $\rho _i $ depends on the specific model, it turns out useful to expand $\rho _i$ as a series in $\Delta \rho = (\rho _L-\rho _R)/2 $ as
\begin{align}
\rho _i = \bar{\rho} + \left(1-\frac{2i}{N} \right) ~\Delta \rho +\sum _{k=2}^{\infty} \mathcal{L}_k(i, \bar{\rho})~\Delta \rho ^k , \label{supp-general-bd-eq-n2}
\end{align}
where $\bar{\rho} = (\rho _L+\rho _R)/2$. For computational convenience, we have written the $k=1$ term separately in the above expression and taken it to be $\mathcal{L}_1(i, \bar{\rho}) = \left(1-\frac{2i}{N} \right)$. This form is sensible because the average density should remain invariant under the transformation $ \rho _L \leftrightarrow \rho _R$ and $i \to (N-i)$. By the same symmetry argument, we must also have $\mathcal{L}_k(i, \bar{\rho})  = (-1)^k\mathcal{L}_k(N-i, \bar{\rho}) $ for all values of $k$. Apart from this symmetry, we do not make any assumption on $\mathcal{L}_k(i, \bar{\rho})$ and their specific forms will depend on the model. 
%For example, in the SSEP and IRW models, we have $\mathcal{L}_2(x, \rho _L) = 0$, whereas in the ZRP model, it is given by $\mathcal{L}_2(x, \rho ) =- x(1-x)/(1+\rho _L) $.

Using the expansion in Eq.~\eqref{supp-general-bd-eq-n2}, the local chemical potential $\mu _i$ and the free energy in Eq.~\eqref{supp-general-bd-eq-2} can also be expanded in $\Delta \rho $ as
\begin{align}
& \mu _i = \bar{\mu} + \sum_{k=1}^{\infty} \frac{a_{k+1}(\bar{\rho})}{k!}  ~\left[\mathcal{K}(i,\bar{\rho}) \right]^k, \\
&  \mathcal{G}_{\mu _i}\left(g \right) = \mathcal{G}_{\bar{\mu} }(g) - g \sum _{k=1}^{\infty} \frac{a_{k+1}(\bar{\rho})}{k!} \left[ \mathcal{K}(i,\bar{\rho}) \right]^k,  \\
& \mathcal{G}_{\mu _i}\left( \rho _i \right) = \mathcal{G}_{\bar{\mu } }(\bar{\rho}) - \bar{\rho} \sum _{k=1}^{\infty} \frac{a_{k+1}(\bar{\rho})}{k!} \left[ \mathcal{K}(i, \bar{\rho}) \right]^k-\sum _{k=2}^{\infty} \frac{(k-1)a_k(\bar{\rho})}{k!} \left[ \mathcal{K}(i, \bar{\rho}) \right]^k,
\end{align}
where we use the notation
\begin{align}
\bar{\mu}  =\frac{\partial a(\bar{\rho})}{\partial \bar{\rho}} ,~~~~ a_k(\bar{\rho}) = \frac{d^k a(\bar{\rho})}{d \bar{\rho} ^k},~~~\mathcal{K}(i,\bar{\rho} ) = \rho _i-\bar{\rho}. \label{kappa-supp}
\end{align}
Plugging the expansions in Eq.~\eqref{supp-general-bd-eq-3} yields
\begin{align}
P\left(n|i\right) = P_{\text{eq}}(n)~\exp \left[ \frac{v_c}{k_B T}  (n/v_c-\bar{\rho}) \sum _{k=1}^{\infty} \frac{a_{k+1}(\bar{\rho})}{k!} \left[ \mathcal{K}(i, \bar{\rho}) \right]^k-\frac{v_c}{k_B T} \sum _{k=2}^{\infty} \frac{(k-1)a_k(\bar{\rho})}{k!} \left[ \mathcal{K}(i, \bar{\rho}) \right]^k  \right], \label{supp-expansion-Pgx}
\end{align}
where $P_{\text{eq}}(n) \sim e^{-\frac{v_c}{k_B T}\left( \mathcal{G}_{\bar{\mu}}\left( n/v_c\right)-\mathcal{G}_{\bar{\mu}}( \bar{\rho} ) \right)}$ is the equilibrium measure with average density $\bar{\rho}$ and chemical potential $\bar{\mu }$. The idea now is to use this series expansion in Eq.~\eqref{appen-avg-information-newnvt} to obtain a perturbation expansion for the positional information. Let us calculate the first term in the expansion.
\subsection{First term}
\noindent
\pss{It turns out, as also demonstrated later, that the first term is of the order $\sim \Delta \rho ^2$. Hence, we include all terms up to this order in the density expansion in Eq.~\eqref{supp-general-bd-eq-n2} and truncate $\mathcal{K}(i, \bar{\rho})$ as
\begin{align}
\mathcal{K}(i, \bar{\rho}) \simeq \mathcal{L}_1(i,\bar{\rho})~\Delta \rho + \mathcal{L}_2(i, \bar{\rho})~\Delta \rho ^2,~~~\text{with } \mathcal{L}_1(i,\bar{\rho})= \left(1-\frac{2i}{N}  \right)
\end{align}
From Eq.~\eqref{supp-expansion-Pgx}, it then follows
\begin{align}
    P(n|i) \simeq P_{\text{eq}}(n)\left[1+f_1(i,n) \Delta \rho+f_2(i,n)\Delta \rho^2\right], \label{abigab}
\end{align}
with two functions $f_1(i,n)$ and $f_2(i,n)$ defined as
\begin{align}
f_1(i,n) &= \frac{v_c }{k_B T}~\mathcal{L}_1(i,\bar{\rho}) ~a_2(\bar{\rho}) \left( n/v_c-\bar{\rho}\right), \\
f_2(i,n) &= \frac{v_c}{2k_B T}\Big\{ -a_2(\bar{\rho})~\mathcal{L}_1(i,\bar{\rho})^2  +  \frac{v_c}{k_B T}a_2(\bar{\rho})^2 \mathcal{L}_1(i,\bar{\rho})^2 (n/v_c-\bar{\rho})^2
 + \Big(  2 a_2(\bar{\rho}) ~\mathcal{L}_2(x, \bar{\rho}) +a_3(\bar{\rho})~\mathcal{L}_1(i,\bar{\rho})^2 \Big)  (n/v_c-\bar{\rho})  \Big\}.        
\end{align}
For a given prior $P_i(i)$, the joint distribution $P(i,n)$ can be written using Bayes' theorem as $P(i,n) = P(n|i) P_i(i)$. Moreover, the marginal distribution $P_n(n)$ can be calculated to be
\begin{eqnarray}
    P_n(n)&=& \sum _{i=1}^{N} P(n|i)~P_i(i),\\
    &\simeq &P_{\text{eq}}(n)\left[1+\left\langle f_1(i,n)\right\rangle_i \Delta \rho+\left\langle f_2(i,n)\right\rangle_i \Delta \rho^2\right],
\end{eqnarray}
with the notation $\langle  f_1(i,n) \rangle _i = \sum _{i=1}^{N}~P_i(i)~f_1(i,n)$ and same for $ \langle f_2(i,n)\rangle_i $. We now have all quantities required to compute the positional information in Eq.~\eqref{appen-avg-information-newnvt}. Inserting these distributions, the first term in $\mathcal{I} _{\left( n, i\right) } \left( \bar{\rho}, \Delta \rho \right)$ can be written as
\begin{align}
\mathcal{I} _{\left( n, i\right) } \left( \bar{\rho}, \Delta \rho \right)  & \approx~ \frac{2v_c^2\Delta \rho ^2~a_2(\bar{\rho})^2 ~\sigma _2(\bar{\rho})}{ (k_B T)^2\ln2} \left[ \Bigg\langle \left( \frac{i}{N}\right)^2  \Bigg\rangle _i-\Bigg\langle \left( \frac{i}{N}\right)\Bigg\rangle_i^2 \right],
 \\
& \approx ~ \frac{2\Delta \rho ^2}{ \ln2~\sigma _2(\bar{\rho})} \left[ \Bigg\langle \left( \frac{i}{N}\right)^2  \Bigg\rangle _i-\Bigg\langle \left( \frac{i}{N}\right)\Bigg\rangle_i^2 \right], \label{supp-general-bd-eq-5}
\end{align}
where $\sigma _k(\bar{\rho})=\langle(n/v_c-\bar{\rho})^k \rangle_{\text{eq}}$ denotes the $k$-th central moment. In writing Eq.~\eqref{supp-general-bd-eq-5}, we have used the fluctuation-response relation $a_2(\bar{\rho}) = k_B T/ \sigma_2(\bar{\rho}) v_c$. For completeness, the proof of this relation is provided in Section~\ref{appen-relations}. To sum up, we have derived the first term in the series expansion of $\mathcal{I} _{\left( n, i\right) } \left( \bar{\rho}, \Delta \rho \right)$ for any prior in terms of the second central moment of the density at equilibrium.} 
\subsection{Second term}
\noindent
In order to obtain the second order term in the expansion, we have to consider higher order terms in the expansion of density in Eq.~\eqref{supp-general-bd-eq-n2}. The second term is of the order $\sim \Delta \rho ^4$ and we therefore take
\begin{align}
\mathcal{K}(i, \bar{\rho} ) \simeq \left(1-\frac{2i}{N} \right)~\Delta \rho + \mathcal{L}_2(i, \bar{\rho})~\Delta \rho ^2+\mathcal{L}_3(i, \bar{\rho})~\Delta \rho ^3 +\mathcal{L}_4(i, \bar{\rho})~\Delta \rho ^4.
\end{align}
We now proceed exactly as before but keeping terms up to order $\sim \Delta \rho ^4$ in the analysis. \pss{This gives us the higher order terms for $\mathcal{I} _{\left( n, i\right) } \left( \bar{\rho}, \Delta \rho \right) $. For the case of flat prior $P_i(i) = 1/N$, the second order term is obtained to be} 
%\begin{align}
%\mathcal{I} _{\left( n, x\right) } \left( \bar{\rho}, \Delta \rho \right) \simeq &~ \frac{\Delta \rho ^2}{6 \ln2~\sigma _2(\bar{\rho})} +  
%  ~\frac{8~v_c~\Delta \rho ^4}{45k_B T \ln 2} \left[ \frac{4 a_2( \bar{\rho}) a_4( \bar{\rho})-5\Big(a_3( \bar{\rho})^2+a_2( \bar{\rho})^3 \Big)}{32 a_2( \bar{\rho})} +\frac{45  a_3( \bar{\rho})}{16 } \int _{0}^{1} dx~\mathcal{L}_2 (x, \bar{\rho })~\big( 4x^2-1 \big) \right.  \nonumber \\
%& \left.      -\frac{45 }{16 } a_2( \bar{\rho}) \left\{ \left( \int _{0}^{1} dx~\mathcal{L}_2 (x, \bar{\rho}) \right)^2-\int _{0}^{1} dx~\mathcal{L}_2 (x, \bar{\rho})^2   -2\int _{0}^{1} dx~\mathcal{L}_3 (x, \bar{\rho})    \big( 1-2x \big) \right\} \right]. 
%\end{align}
\begin{align}
\mathcal{I} _{\left( n, i\right) } \left( \bar{\rho}, \Delta \rho \right) \simeq &~ \frac{\Delta \rho ^2}{6 \ln2~\sigma _2(\bar{\rho})} +  
  ~\frac{8~v_c~\Delta \rho ^4}{45k_B T \ln 2} \left[ \frac{4 a_2( \bar{\rho}) a_4( \bar{\rho})-5\Big(a_3( \bar{\rho})^2+a_2( \bar{\rho})^3 \Big)}{32 a_2( \bar{\rho})} +\frac{45  a_3( \bar{\rho})}{16 N} \sum _{i=1}^{N}~\mathcal{L}_2 (i, \bar{\rho })~\left( \frac{4i^2}{N^2}-1 \right) \right.  \nonumber \\
& \left.      -\frac{45 }{16 } a_2( \bar{\rho}) \left\{ \left( N^{-1} \sum _{i=1}^N~\mathcal{L}_2 (i, \bar{\rho}) \right)^2-N^{-1} \sum _{i=1}^N~\mathcal{L}_2 (i, \bar{\rho})^2   -2N^{-1} \sum _{i=1}^N~\mathcal{L}_3 (i, \bar{\rho})    \mathcal{L}_1 (i, \bar{\rho})  \right\} \right]. 
\end{align}
For later comparison, it is useful to write this expression in terms of the $\sigma _k(\bar{\rho})$. To achieve this, we use the following set of relations
\begin{align}
\frac{v_c a_2(\bar{\rho}) }{k_B T}= \frac{1}{ \sigma_2(\bar{\rho})},~~~ \frac{v_c a_3(\bar{\rho})}{k_B T} = -\frac{\sigma_3(\bar{\rho})}{\sigma_2(\bar{\rho})^3},~~\frac{v_C a_4(\bar{\rho})}{k_B T} = -\frac{\sigma_4(\bar{\rho})}{\sigma_2(\bar{\rho})^4}+\frac{3}{\sigma_2(\bar{\rho})^5} \big[ \sigma_3(\bar{\rho})^2+\sigma_2(\bar{\rho})^3\big], \label{FTT-appen}
\end{align}
which have been derived in Section~\ref{appen-relations}. The above expression now becomes
\begin{align}
\mathcal{I} _{\left( n, i\right) } \left( \bar{\rho}, \Delta \rho \right) \approx &~ \frac{\Delta \rho ^2}{6 \ln2~\sigma _2(\bar{\rho})} +  
  ~\frac{8~\Delta \rho ^4}{45 \ln 2~\sigma _2(\bar{\rho})} \left[ \frac{7\Big(\sigma_3( \bar{\rho})^2+\sigma_2( \bar{\rho})^3 \Big)-4 \sigma_2( \bar{\rho}) \sigma_4( \bar{\rho})}{32 ~\sigma_2( \bar{\rho})^4} +\frac{45 \sigma_3( \bar{\rho})}{16 \sigma _2(\bar{\rho})^2 N} \sum _{i=1}^{N}~\mathcal{L}_2 (i, \bar{\rho })~\left( \frac{4i^2}{N^2}-1 \right)\right.  \nonumber \\
& \left.      -\frac{45}{16} \left\{ \left( N^{-1} \sum _{i=1}^N~\mathcal{L}_2 (i, \bar{\rho}) \right)^2-N^{-1} \sum _{i=1}^N~\mathcal{L}_2 (i, \bar{\rho})^2   -2N^{-1} \sum _{i=1}^N~\mathcal{L}_3 (i, \bar{\rho})    \mathcal{L}_1 (i, \bar{\rho})  \right\} \right]. \label{appen-exp-2terms}
\end{align}
This gives the first two terms in the series expansion of the positional information. One can also obtain the higher order terms in the same way. In summary, we have developed a perturbative approach to calculate $\mathcal{I} _{\left( n, i\right) } \left( \bar{\rho}, \Delta \rho \right) $ for a general boundary-driven system. Our approach relies on the local equilibrium assumption and requires only the knowledge of the average density $\rho _i$. Positional information is then obtained as an expansion in $\Delta \rho$ with coefficients depending on the equilibrium central moments of the density (or equivalently the derivatives of the free energy). 
\subsection{Connection of $\mathcal{I} _{\left( n, i\right) } \left( \bar{\rho} , \Delta \rho \right)$ with the chemical potential difference \pss{for a general $P_i(i)$}}
\noindent
Having developed a methodology to compute positional information, we are now in a position to establish its connection with the chemical potential difference driving the system, $\Delta \mu = \mu _L-\mu_R$. Following Eq.~\eqref{supp-general-bd-eq-2}, this can be written as
\begin{align}
\Delta \mu & = a_1(\bar{\rho } +\Delta \rho)-a_1(\bar{\rho } -\Delta \rho)  \simeq \frac{2k_B T \Delta \rho }{v_c\sigma _2(\bar{\rho} )} .  \label{kmayq}
\end{align}	
Combining this with the expression of $\mathcal{I} _{\left( n, i\right) } \left( \bar{\rho}, \Delta \rho \right) $ in Eq.~\eqref{supp-general-bd-eq-5}, we obtain for the leading order in $\Delta \rho$
%\begin{align}
%\mathcal{I} _{\left( n, i\right) } \left( \bar{\rho} , \Delta \rho \right) \simeq \frac{\Delta \mu ~\Delta \rho }{ 12 \ln 2}. \label{old-appen-pj8q} 
%\end{align}
%Reinstating the $k_B T$ term, the expression becomes
\pss{\begin{align}
\mathcal{I} _{\left( n, i\right) } \left( \bar{\rho} , \Delta \rho \right) \approx \frac{\Delta \mu ~v_c ~ \Delta \rho }{ k_B T \ln 2}\left[ \Bigg\langle \left( \frac{i}{N}\right)^2  \Bigg\rangle _i-\Bigg\langle \left( \frac{i}{N}\right)\Bigg\rangle_i^2 \right]. \label{appen-pj8q}
\end{align}}
This relation quantitatively gives the link between the positional information and non-equilibrium nature of the system for a given prior. At least with only the first term in $\mathcal{I} _{\left( n, i\right) } \left( \bar{\rho} , \Delta \rho \right) $ and $\Delta \mu$, Eq.~\eqref{appen-pj8q} tells us how positional information increases on increasing the non-equilibrium drive. \\
\indent
\pss{We are now interested in optimising Eq.~\eqref{appen-pj8q} with respect to $P_i(i)$. First recall that the index $i$ can take only positive integer values and is bounded as $1 \leq i \leq N$. This means
\begin{align}
    \frac{i}{N}\leq 1 \implies \left( \frac{i}{N}\right)^2 \leq \left( \frac{i}{N}\right).
\end{align}
Averaging both sides
\begin{align}
    \Bigg\langle \left( \frac{i}{N}\right)^2 \Bigg\rangle_i \leq \Bigg\langle \frac{i}{N} \Bigg\rangle_i
\end{align}
Using this in Eq.~\eqref{appen-pj8q}, one can write the bound
\begin{align}
    \mathcal{I} _{\left( n, i\right) } \left( \bar{\rho}, \Delta \rho \right) \leq \frac{\Delta \mu ~v_c~\Delta \rho }{ k_B T\ln2} \left[ \Bigg\langle  \frac{i}{N}  \Bigg\rangle_i-\Bigg\langle  \frac{i}{N} \Bigg\rangle_i^2 \right]. \label{appen-non-falt-ps-213}
\end{align}
Let us find the bound on the term inside $[..]$. Using the following relation
\begin{align}
    & \left( \Bigg\langle\frac{i}{N} \Bigg\rangle_i-\frac{1}{2}  \right)^2 \geq 0, \implies  \left[ \Bigg\langle  \frac{i}{N}  \Bigg\rangle_i-\Bigg\langle  \frac{i}{N} \Bigg\rangle_i^2 \right] \leq \frac{1}{4}
\end{align}
in the above inequality, we obtain
\begin{align}
  \mathcal{I} _{\left( n, i\right) } \left( \bar{\rho}, \Delta \rho \right) \leq \frac{\Delta \mu ~v_c~\Delta \rho }{ 4 k_B T\ln2} \label{appen-non-falt-ps-214}. 
\end{align}
This gives a universal bound to the near-equilibrium positional information that is valid for for any boundary-driven systems with arbitrary choice of prior.\\
\indent
Beyond linear response regime, writing any universal relation is difficult due to the dependence of higher-order terms in positional information on the specific model, as illustrated in Eq.~\eqref{appen-exp-2terms}. In absence of a general formulation, we have studied several solvable boundary-driven systems. For each of these models, we find that the upper bound~\eqref{appen-non-falt-ps-214} is valid even in the far-from equilibrium conditions and we obtain a fundamental limit on the positional information in terms of the system's distance from the equilibrium.\\
\indent
For the case of a flat prior, our analysis on these solvable models reveal that there exists a tighter bound to the positional information compared to the one in \eqref{appen-non-falt-ps-214}
\pss{\begin{align}
\mathcal{I} _{\left( n, i\right) } \left( \bar{\rho} , \Delta \rho \right)  \leq \frac{\Delta \mu ~v_c~\Delta \rho }{ 12~k_B T \ln 2}. \label{appen-ineqq-eq-1}
\end{align}
This bound is saturated if the system is close to equilibrium. This can be verified by plugging the variance
\begin{align}
\left[ \Bigg\langle \left( \frac{i}{N}\right)^2  \Bigg\rangle _i-\Bigg\langle  \frac{i}{N}\Bigg\rangle_i^2 \right] \simeq \frac{1}{12},~~\text{for large }N
\end{align}
in our general expression in Eq.~\eqref{appen-pj8q}. In what follows, we will first look at the flat prior case and demonstrate numerically the upper bound~\eqref{appen-ineqq-eq-1} for different toy models when they are arbitrarily from from the  equilibrium. After that, we will investigate how positional information varies for different priors in section~\ref{non-flat} and test the other bound~\eqref{appen-non-falt-ps-214} for these models. Below, we first look at SSEP followed by a number of other models. \pss{In rest of our analysis, we will set $k_B T=1$ and $v_c=1$ for simplicity.}}}

\section{Positional information for the boundary-driven SSEP}
\label{apen-sep-caba12}
One of the models that we discussed in the main text is the open symmetric simple exclusion process (SSEP). The model consists of $N$ lattice sites in one dimension represented by the index $i$ that runs from $1$ to $N$. Each lattice site can either be vacant $(n_i =0)$ or occupied by a single particle $(n_i =1)$. The dynamics of the particles are as follows: Within the bulk, where $1 < i < N$, a particle can jump to either of its neighbouring sites with a rate $p$, provided that these neighbouring sites are empty. On the other hand, the two boundary sites ($i=1$ and $i=N$) are in contact with two different particle reservoirs, and their dynamics are modified accordingly.  At $i=1$, a particle can be added (removed) with a rate $\alpha_L$ ($\beta_L$) if the site is vacant (occupied). Conversely, at $i=N$, a particle can be added (removed) with a rate $\alpha_R$ ($\beta_R$) if it is vacant (occupied). At any small time interval $[t, t+dt]$, the occupancy variable $n_i(t)$ for each lattice site evolves according to the following update rule \cite{PhysRevLett.87.150601, Derrida2002}
%Let us first consider the interacting case consisting of one dimensional lattice with $N$ sites. For all $1 <i<N$, particle can hop with probability $pdt$ on either sides provided that its neighbouring site is empty. The dynamics at the end sites (namely $i=1$ and $i=N$) is different. For $i=1$, a particle can be added (removed) with probability $\alpha _L dt$  ($\beta _L dt$) if the site if filled (vacant). On the other hand, for $i=N$, a particle can be added (removed) with probability $\alpha _R dt$  ($\beta _R dt$) if it is filled (vacant). Furthermore, we assume that $\alpha _L > \alpha _R$ and $\beta _L < \beta_R$ so that the we have a reservoir at the $i=1$ and sink at $i=N$.
%\begin{figure}[!htbp]
%	\includegraphics[scale=0.6]{model}
%%	\includegraphics[scale=0.4]{heat-dist.jpeg}
%	\centering
%	\caption{Boundary driven SSEP} 
%	\label{model-fig}
%\end{figure}
%For each lattice site, we assign a variable $n_i(t) \in \{0,1 \}$ where $n_i(t)=1$ if it has a particle at that instant and $n_i(t)=0$ otherwise. The update rule for each lattice site can be written as
\begin{align}
& n_i(t + \Delta t) - n _i(t)~ = \begin{cases}
n_{i+1}(t),~~~~~~~~~~~~\text{w.p.} \left[1-n_i(t) \right]~p \Delta t, \\
n_{i-1}(t),~~~~~~~~~~~~\text{w.p.} \left[1-n_i(t) \right]~p \Delta t,\\
-n_{i}(t),~~~~~~~~~~~~~\text{w.p.} \left[1-n_{i+1}(t) \right]~p \Delta t,  \\
-n_{i}(t),~~~~~~~~~~~~~\text{w.p.} \left[1-n_{i-1}(t) \right]~p \Delta t, \\
~~~0,~~~~~~~~~~~~~~~~~\text{otherwise},
\end{cases},~~~~~~~\text{for }i\neq 1,~~i \neq N \\
& n_1(t + \Delta t) - n _1(t) ~= \begin{cases}
-n_1(t),~~~~~~~~~~~~\text{w.p.} \left[1-n_2(t) \right]~p \Delta t ,\\
~~n_2(t),~~~~~~~~~~~~\text{w.p.} \left[1-n_1(t) \right]~p \Delta t ,\\
~~~1,~~~~~~~~~~~~~~~~\text{w.p.} \left[1-n_1(t) \right]~\alpha _L \Delta t ,\\
-1,~~~~~~~~~~~~~~~~~\text{w.p.} ~~~n_1(t)~\beta _L \Delta t, \\
~~0,~~~~~~~~~~~~~~~~~~\text{otherwise},
\end{cases}\\
& n_N(t + \Delta t) - n _N(t) = \begin{cases}
-n_N(t),~~~~~~~~~~~\text{w.p.} \left[1-n_{N-1}(t) \right]~p \Delta t ,\\
~~n_{N-1}(t),~~~~~~~~\text{w.p.} \left[1-n_N(t) \right]~p \Delta t, \\
~~~1,~~~~~~~~~~~~~~~~\text{w.p.} \left[1-n_N(t) \right]~\alpha _R \Delta t, \\
-1,~~~~~~~~~~~~~~~~~\text{w.p.} ~~~n_N(t)~\beta _R \Delta t, \\
~~0,~~~~~~~~~~~~~~~~~\text{otherwise},
\end{cases}
\end{align}
where $``\text{w.p}"$ is the short-hand notation for ``with probability". Our aim is to calculate the positional information for this model. As evident from its definition, it is then necessary to compute the conditional probability $P(n|i )$ of having $n$ particles at $i$-th site. Later, we show that this probability can be expressed in terms of the mean density $\rho _i(t) = \langle n_i(t) \rangle$. Therefore, in what follows, we will first calculate $\rho _i(t)$.

\subsection{Average density $\rho _i(t) = \langle n_i(t) \rangle$}
From the update rules written above, one can write the differential equations for $\rho _i(t)$ as
\begin{align}
& \frac{\partial \rho _i(t)}{ \partial t}  = p \left[ \rho _{i+1}(t)+\rho _{i-1}(t) -2 \rho _{i}(t)\right],~~~~~~~(\text{for }i\neq 1,~~i \neq N) \\
& \frac{\partial \rho _1(t)}{ \partial t}  = \alpha _L - (p+\alpha _L+\beta _L) \rho _1(t)+p ~\rho _2(t), \\
& \frac{\partial \rho _N(t)}{ \partial t}  = \alpha _R - (p+\alpha _R+\beta _R) \rho _1(t)+p ~\rho _{N-1}(t).
\end{align}
Since we are interested in the steady-state properties, we replace the time derivatives on the left hand sides of these equations by zero and obtain the solution as
%One can solve these equations to show that the density in the steady state is
\begin{align}
&~\rho _i (t \to \infty)  = \mathcal{A} + (i-1) ~\mathcal{B}, \label{avg-density-eq} \\
& ~\text{with }\mathcal{A} =\frac{\alpha _L \left( \alpha_R + \beta _R  \right)(N-1) +p \left( \alpha _L + \alpha _R \right)   }{\left( \alpha_L + \beta _L  \right) \left( \alpha_R + \beta _R  \right) (N-1)+p \left( \alpha_L + \beta _L+\alpha_R + \beta _R  \right)}, \\
&~~~~~~~~\mathcal{B}  = \frac{\alpha _R \beta _L-\alpha _L \beta _R}{\left( \alpha_L + \beta _L  \right) \left( \alpha_R + \beta _R  \right) (N-1)} .
\end{align}
%Meanwhile, due to two reservoirs at the ends, the system will also have a non-zero current $J_i(t)$ which measures number of particles jumped from $i \to (i+1)$ minus the number of particles jumped from $(i+1) \to i$. At steady state, we have
%\begin{align}
%J_i &= \langle n_i(1-n_{i+1})   \rangle-\langle n_{i+1}(1-n_{i})  \rangle , \\
%& = \rho _i -\rho _{i+1}, \\
%& = -\mathcal{B}. \label{avg-current-eq}
%\end{align}
%For large $N$, we introduce a rescaled (continuous) variable $x = i/N$ where $x \in [0,1]$ and rewrite $\rho(x) = \rho _i(t \to \infty)$
For large $N$, one can rewrite this expression as
\begin{align}
\rho_i& \simeq \rho _L - \left( \rho _L- \rho _R \right) \frac{i}{N}, ~~~\text{with }\rho _L = \frac{\alpha _L}{\alpha _L + \beta _L},~~\rho _R = \frac{\alpha _R}{\alpha _R + \beta _R}. \label{ackbxq-new}
\end{align}
We have used $\rho _L$ and $\rho _R$ to represent the average densities of the left and the right reservoir respectively and are expressed in terms of the parameters of our model. Furthermore, the approximate equality in Eq.~\eqref{ackbxq} is used to indicate that we are working in the large $N$ limit. Finally, we express density in terms of the variables $\bar{\rho} = (\rho _L + \rho _R)/2$ and $\Delta \rho = (\rho _L - \rho _R)/2$
\begin{align}
\rho_i\simeq ~\bar{\rho}+\left(1- \frac{2i}{N} \right)~ \Delta \rho, \label{ackbxq}
\end{align}
and utilize this form to derive the positional information.
%\begin{figure}[t]
%\centering
%  \includegraphics[scale=0.4]{Inform-SSEP.pdf}
 % %\includegraphics[scale=0.3]{dist-tmax-BM-new.pdf}
 % \caption{(a) Non-monotonic nature of the positional information as a function of the density $\rho _L$ for different values of $\Delta \rho$ for the open SSEP.}
 % \label{SM-avginf-fig}
%\end{figure}
\begin{figure}[t]
\centering
  \includegraphics[scale=0.5]{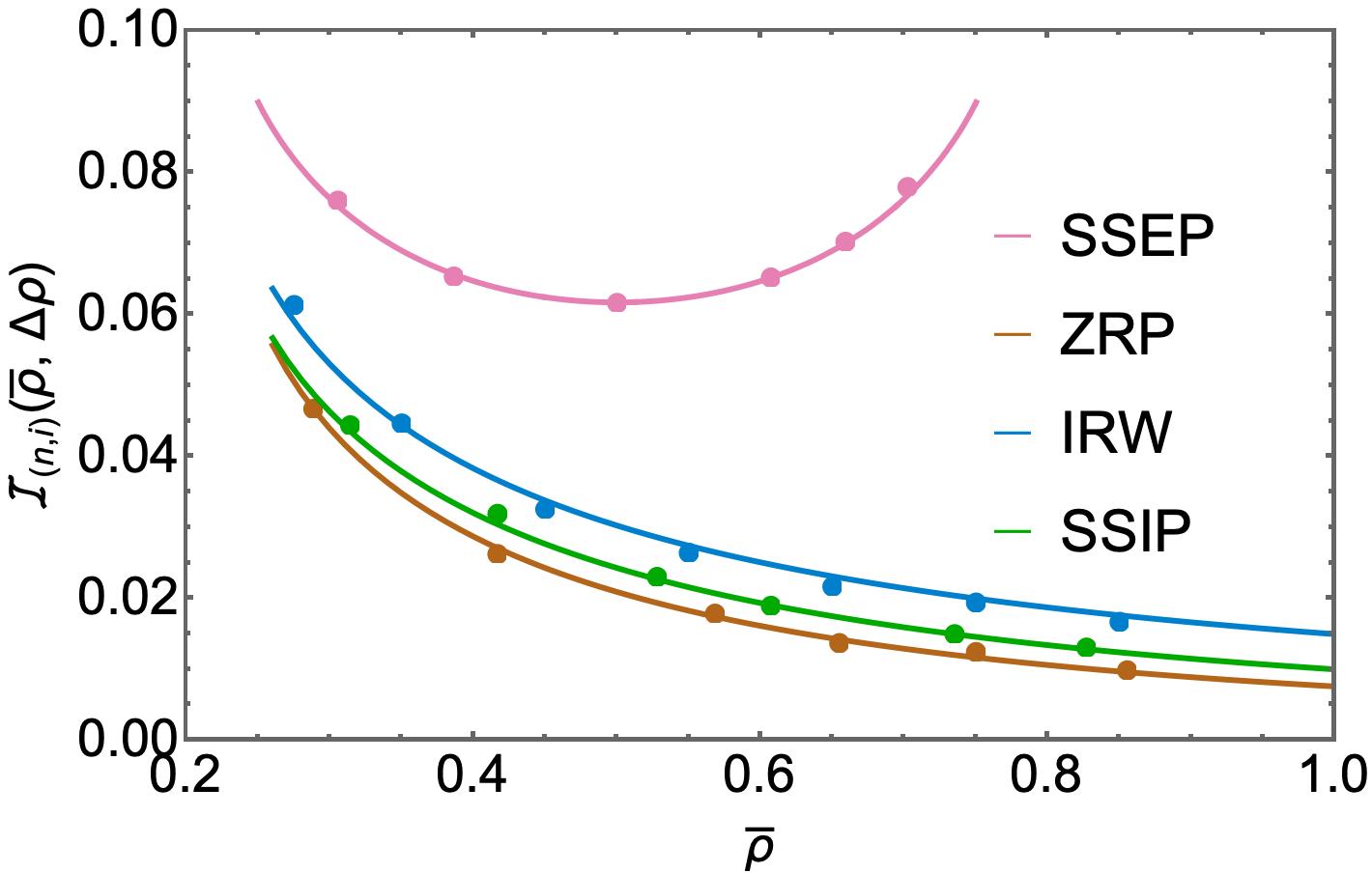}
  \caption{Plot of the positional information as a function of $\bar{\rho}$ for four different boundary-driven processes and its comparison with the numerical simulations.  In each case, solid lines illustrate the analytical expression, while symbols denote simulation data. We have fixed the bias to $\Delta \rho = 0.25$ and chosen the flat prior.}
  \label{SM-avginf-fig}
\end{figure}
\subsection{Positional information} 
\noindent
In order to obtain the positional information, we will use its definition in Eq.~\eqref{appen-avg-information} and calculate the entropies $S[P(n|i )]$ and $S[P_n(n)]$. Recall that $n_i$ is a binary variable that can take values either $0$ or $1$ depending on whether the site is vacant or occupied. We therefore have $\rho_i = 1 \times P(n=1|i) +0 \times P(n=0|i) =  P(n=1|i) $. The complementary probability will then simply be $P(n=0|i)=1-\rho_i$. 
\begin{align}
P(n=1|i) &= \rho_i = \bar{\rho}+\left(1- \frac{2i}{N} \right)~ \Delta \rho,  \label{appen-cond-ssep-1}\\
P(n=0|i) &=1-\rho _i=1-\bar{\rho}-\left(1- \frac{2i}{N} \right)~ \Delta \rho .\label{appen-cond-ssep-2}
\end{align}
%\begin{align}
%& P(\sigma =1|x)  = P(n=1|x) = \langle n(x) \rangle = \rho _L - (\rho _L - \rho _R)x, \\
%& P(\sigma =-1|x) = 1-P(\sigma =1|x) = 1-\rho _L + (\rho _L - \rho _R)x.
%\end{align}
From these, we get
\begin{align}
S[P(n|i )] & = - \sum _{n =\{0,1 \} } P(n |i) \log _2 [P(n |i)], \\
& = -\left[\bar{\rho}+\left(1- \frac{2i}{N} \right)~ \Delta \rho\right] \log _2\left[ \bar{\rho}+\left(1- \frac{2i}{N} \right)~ \Delta \rho \right] \nonumber \\
& ~~~~~~~~~~~~~~-\left[1-\bar{\rho}-\left(1- \frac{2i}{N} \right)~ \Delta \rho \right] \log _2\left[ 1-\bar{\rho}-\left(1- \frac{2i}{N} \right)~ \Delta \rho \right].
\end{align}
For the average information in Eq.~\eqref{appen-avg-information}, we also have to perform averaging of $S[P(n|i )]$ over the prior distribution $P_i(i)$ which we have chosen to be uniform. This yields
%perform averaging over the position $x$. In absence of any morphogen signal, we assume that all cells are equally which gives $P_x(x) = 1$. Therefore, we get
\begin{equation}
\scalebox{1.0}{$
\begin{split}
\langle  S[P(n|i )]  \rangle _i &= \sum_{i=1}^{\infty}~ P_i(i)~S[P(n|i )] =\frac{1}{N} \sum_{i=1}^{\infty}~S[P(n|i )] 
\end{split}$}
 \nonumber
\end{equation}
Performing this summation turns out to be difficult. We therefore introduce a change of variable $x = i/N$ and take $N \to \infty$. One can then use the transformation $1/N \sum _{i=1}^{N} \to \int_0^1~ dx$ and obtain
\begin{equation}
\scalebox{1.0}{$
\begin{split}
\langle  S[P(n|i )]  \rangle _i & = -\int _{0}^{1} dx \Bigg[\left[\bar{\rho}+(1-2x) \Delta \rho \right] \log _2\left[ \bar{\rho}+(1-2x)\Delta \rho \right]+\left[1-\bar{\rho}-(1-2x) \Delta \rho \right] \log _2\left[ 1-\bar{\rho}-(1-2x) \Delta \rho \right] \Bigg], \\
& = -\frac{ 1}{ \ln 16 ~(\Delta \rho)} \Big[ \left( \bar{\rho}+\Delta \rho \right) ^2 \ln \left( \bar{\rho}+\Delta \rho \right) - \left( \bar{\rho}-\Delta \rho \right)^2 \ln \left( \bar{\rho}-\Delta \rho \right) -2\Delta \rho +\left(1- \bar{\rho} +\Delta \rho \right)^2 \ln \left( 1-\bar{\rho }+\Delta \rho \right) \Big. \\
& ~~~~~~~~~~~~~~~~~~~~~~~\Big.- \left( 1-\bar{\rho}-\Delta \rho \right)^2 \ln \left( 1-\bar{\rho}-\Delta \rho \right) \Big]. 
\end{split}$}
 \label{entropy-fin}
\end{equation}
We next compute the other entropy, $S[P_n (n)]$, in the definition of $\mathcal{I} _{\left( n, i\right) } \left( \bar{\rho} , \Delta \rho \right) $. We can write $P_n(n)$ by summing over $i$ in the joint probability $P(n,i) = P(n|i)$ and then using Eqs.~\eqref{appen-cond-ssep-1} and \eqref{appen-cond-ssep-2} for $P(n|i)$. This gives
\begin{align}
P_{n}(n) & =\frac{1}{N} \sum _{i=1}^N~ P(n|i) = \bar{\rho} ~\delta _{n ,1}+\left( 1- \bar{\rho} \right)~ \delta _{n ,0}.
\end{align}	
Here $\delta _{n,1}$ is the Kronecker delta which takes value one if $n=1$ and zero otherwise (similarly for $\delta _{n,0}$). Using this expression above, we find
\begin{align}
S[P_n (n)] & = -\sum _{ \{n =0,1 \} } P_{n}(n) \log _2 \left[  P_{n}(n)  \right] = -\bar{\rho} \log _2 \bar{\rho} -(1-\bar{\rho}) \log _2 (1-\bar{\rho}). \label{entropy-ini}
\end{align}
We now have all quantities essential for calculating the positional information. Substituting Eqs.~\eqref{entropy-fin} and \eqref{entropy-ini} in Eq.~\eqref{appen-avg-information}, the final expression is
%\begin{align}
%\mathcal{I} _{\left( n, x\right) } \left( \bar{\rho}, \Delta \rho \right) = & \frac{1}{\ln 16~\Delta \rho} \left[ \left( \bar{\rho} +\Delta \rho \right)^2 \ln \left( \bar{\rho} +\Delta \rho \right)  -   \left( \bar{\rho} -\Delta \rho \right)^2 \ln \left( \bar{\rho} -\Delta \rho \right)  +  \left(1- \bar{\rho} +\Delta \rho \right)^2 \ln \left(1- \bar{\rho} +\Delta \rho \right) \right. \nonumber \\
%& - \left. \left(1- \bar{\rho} -\Delta \rho \right)^2 \ln \left(1- \bar{\rho} -\Delta \rho \right) -2 \Delta \rho \right]- \left( 1-\bar{\rho} \right) \log _2  \left( 1-\bar{\rho} \right) -\bar{\rho }\log _2 \bar{\rho }.  \label{I-int-eqn}
%\end{align}
\begin{align}
\mathcal{I} _{\left( n, i\right) } \left( \bar{\rho}, \Delta \rho \right) = & ~\frac{1}{\ln 16~\Delta \rho} \Big[ \mathcal{Y}(\bar{\rho}, \Delta \rho)+\mathcal{Y}(1-\bar{\rho}, \Delta \rho)  -2 \Delta \rho \Big]- \left( 1-\bar{\rho} \right) \log _2  \left( 1-\bar{\rho} \right) -\bar{\rho }\log _2 \bar{\rho }, \label{I-int-eqn} \\
\text{where } \mathcal{Y} (\bar{\rho}, \Delta \rho) & = \left( \bar{\rho} +\Delta \rho \right)^2 \ln \left( \bar{\rho} +\Delta \rho \right)  -   \left( \bar{\rho} -\Delta \rho \right)^2 \ln \left( \bar{\rho} -\Delta \rho \right). \label{y-fun-eq}
\end{align}
This result has been quoted in the main text. Eq.~\eqref{I-int-eqn} is also consistent with our general result in Eq.~\eqref{appen-exp-2terms} in the $\Delta \rho \to 0$ limit. In Figure~\ref{SM-avginf-fig}, we have plotted $\mathcal{I} _{\left( n, i\right) } \left( \bar{\rho}, \Delta \rho \right)$ as a function of $\bar{\rho}$ and compared it with numerical simulations. We see an excellent agreement between our theoretical formula and the numerics. Notice that for a given $\Delta \rho $, $\bar{\rho}$ can vary between $\Delta \rho$ and  $(1-\Delta \rho)$. Within this range, we find that the positional information for SSEP changes in a non-monotonic manner with $\bar{\rho}$. To understand this heuristically, let us see what happens when $\bar{\rho} \to (1-\Delta \rho)$ or equivalently $\rho _L \to 1$. In terms of the jump rates, this means that  $\alpha _L \gg \beta _L$ and sites near the left boundary are more likely to be occupied with particles than those near the right boundary. Hence, if we pick a site randomly but is occupied by a particle, it is more likely to be closer to the left boundary than the right one. From the probabilistic perspective, this would imply that the distribution $P(i|n=1)$ is sharply peaked at $i=1$, while the complementary probability $P(i|n=0)$ peaks at $i=N$. The resulting entropy $\langle S[P(i|n )] \rangle _{n}$ associated with these peaked distributions, as expressed in Eq.~\eqref{appen-pos-inf-eq-newra}, is small which leads to a larger value of $\mathcal{I} _{\left( n, i\right) } \left( \bar{\rho}, \Delta \rho \right)$. Similarly, for $\bar{\rho } \to \Delta \rho$, one can argue that the probabilities $P(i|n=1)$ and $P(i|n=0)$ are narrow functions of $i$ which again give large $\mathcal{I} _{\left( n, i\right) } \left( \bar{\rho} , \Delta \rho \right)$. In between these large values, $P(i|n=1)$ and $P(i|n=0)$ are broadest at some $\bar{\rho} $ for which the entropy $\langle S[P(i|n )] \rangle _{n}$ is highest. This in turn leads to the smallest positional information via Eq.~\eqref{appen-pos-inf-eq-newra}.
\begin{figure}[t]
	\includegraphics[scale=0.45]{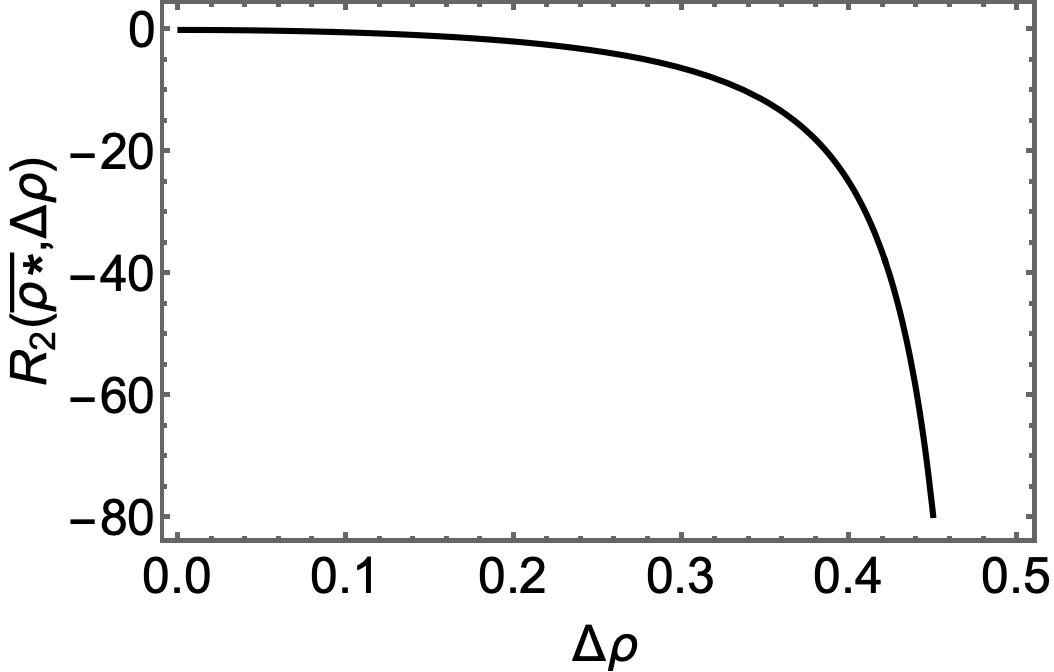}
	\includegraphics[scale=0.38]{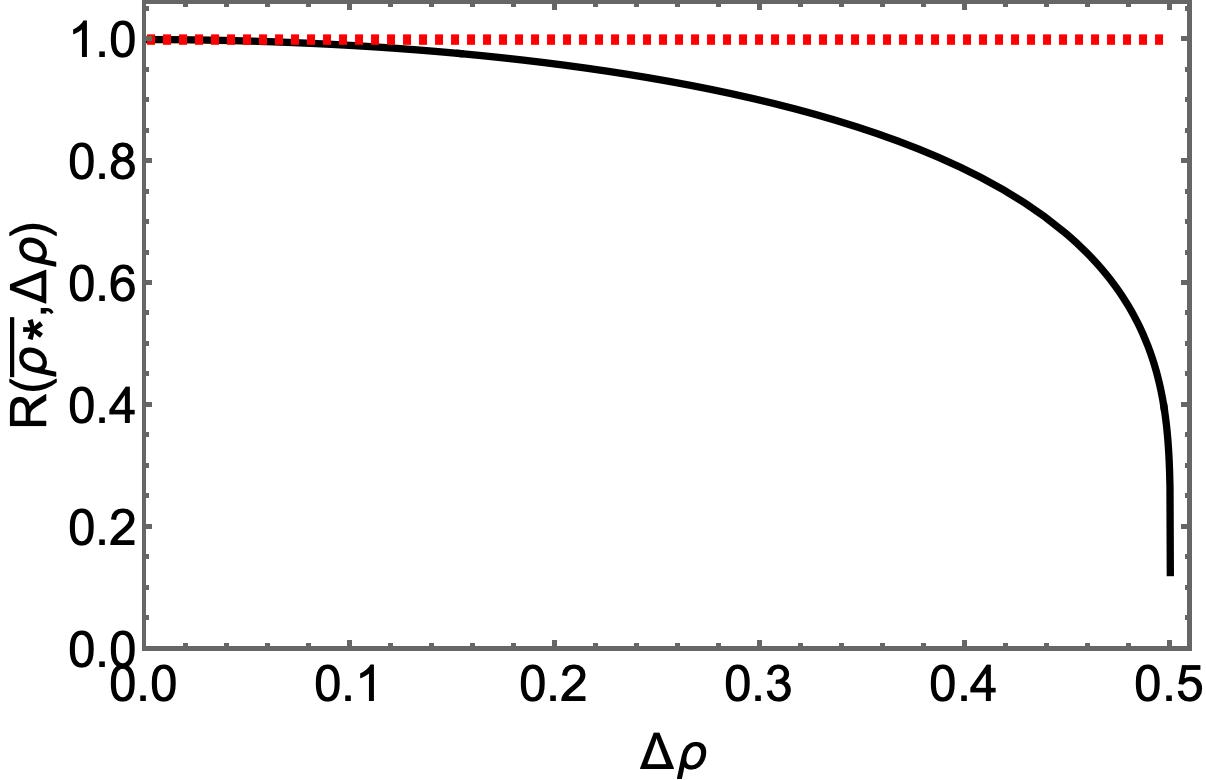}
	\centering
	\caption{\textit{Left panel:} Plot of the second derivative $R_2(\bar{\rho}^*, \Delta \rho) = \frac{\partial^2 R(\bar{\rho}, \Delta \rho)}{\partial \bar{\rho}^2} \Big| _{\bar{\rho} = \bar{\rho}^*}$ where $\bar{\rho}^* = 1/2$. For all $\Delta \rho$, the second derivative takes only negative values. \textit{Right panel:} Illustration of $R(\bar{\rho}^*,\Delta \rho)$ in Eq.~\eqref{Rmin-eq} as a function of $\Delta \rho $. The solid black line shows the plot while red dashed line indicates that $R(\bar{\rho}^*,\Delta \rho)$ is always bounded by the value one.} 
	\label{Rmin-fig}
\end{figure}
\subsection{Upper bound on $\mathcal{I} _{\left( n, i\right) } \left( \bar{\rho} , \Delta \rho \right)$}
\noindent
We will now use the derived expression of $\mathcal{I} _{\left( n, i\right) } \left( \bar{\rho} , \Delta \rho \right)$ to prove the upper bound in \eqref{appen-ineqq-eq-1}. Observe that the chemical potential difference $\Delta \mu$ is expressed in terms of the first derivative of the associated free energy in Eq.~\eqref{kmayq}. To obtain this, we use the fluctuation-response relation in Eq.~\eqref{FTT-appen} as
\begin{align}
a_2(\bar{\rho}) = \frac{1}{\sigma _2(\bar{\rho})} = \frac{1}{\bar{\rho}(1-\bar{\rho})}, \implies a_1(\bar{\rho}) = \ln \left( \frac{\bar{\rho}}{1-\bar{\rho}} \right).
\end{align}
where $\sigma _2(\bar{\rho}) = \bar{\rho}(1-\bar{\rho})$ follows from Eqs.~\eqref{appen-cond-ssep-1} and \eqref{appen-cond-ssep-2} with $\Delta \rho = 0$ \greenw{and $k_BT=1$ is assumed}. Using Eq.~\eqref{kmayq}, we now get
\begin{align}
\Delta \mu = \ln \left( \frac{\bar{\rho}+ \Delta \rho}{1-\bar{\rho}-\Delta \rho }  \right)-\ln \left( \frac{\bar{\rho}- \Delta \rho}{1-\bar{\rho}+\Delta \rho }  \right). \label{safiqp}
\end{align} 
To derive the bound \eqref{appen-ineqq-eq-1}, it is useful to define a ratio $R(\bar{\rho}, \Delta \rho)$ as
\begin{align}
R(\bar{\rho}, \Delta \rho) = \frac{12 \ln 2 ~\mathcal{I} _{\left( n, i\right) }(\bar{\rho}, \Delta \rho)}{\Delta \mu~ \Delta \rho}.
\end{align}
We are interested in finding the maximum value of this ratio. Below we show that this maximum value turns out to be one thereby proving the upper bound. Proceeding ahead, we take the first derivative of $R(\bar{\rho}, \Delta \rho)$ with respect to $\bar{\rho}$ for fixed $\Delta \rho$ 
\begin{align}
\frac{\partial R(\bar{\rho}, \Delta \rho)}{\partial \bar{\rho}}  = R(\bar{\rho}, \Delta \rho) \left[ \frac{1}{\mathcal{I} _{\left( n, i\right) } \left( \bar{\rho}, \Delta \rho \right) }       \frac{\partial~\mathcal{I} _{\left( n, i\right) } \left( \bar{\rho}, \Delta \rho \right)  }{\partial \bar{\rho}}  -\frac{1}{\Delta \mu}\frac{\partial \Delta \mu}{\partial \bar{\rho}} \right],
\end{align}
and setting it to zero, we find the condition for optimality as $\bar{\rho}^* = 1/2$. Moreover, the second derivative of $ R(\bar{\rho}, \Delta \rho)$ at $\bar{\rho}^*$ is always negative indicating that the extremum is a maxima. This is illustrated in the left panel of Figure~\ref{Rmin-fig}. Therefore, we obtain
\begin{align}
R(\bar{\rho}, \Delta \rho) & \leq R \left(\bar{\rho}^*=\frac{1}{2}, \Delta \rho \right) , \nonumber \\
& = \frac{3}{2 \Delta \rho ^2} \frac{        (1+4~ \Delta \rho ^2)~\text{Arctanh}(2 \Delta \rho)-2 \Delta \rho + 2 \Delta \rho~ \ln (1-4 \Delta \rho ^2) }{\ln (1+2 \Delta \rho)-\ln (1-2 \Delta \rho)}. \label{Rmin-eq}
\end{align}
Plotting this expression in Figure~\ref{Rmin-fig} (right panel), we find that the value of $R \left(\bar{\rho}^*, \Delta \rho \right)$ is always upper bounded by one. From our analysis above, this translates to $R(\bar{\rho}, \Delta \rho) \leq 1$ and we finally get
\begin{align}
\mathcal{I} _{\left( n, i\right) } \left( \bar{\rho} , \Delta \rho \right)  \leq \frac{\Delta \mu ~\Delta \rho }{ 12 \ln 2}. \label{mqobxt}
\end{align}
 \begin{figure}[t]
  \centering
  \subfigure{\includegraphics[scale=0.3]{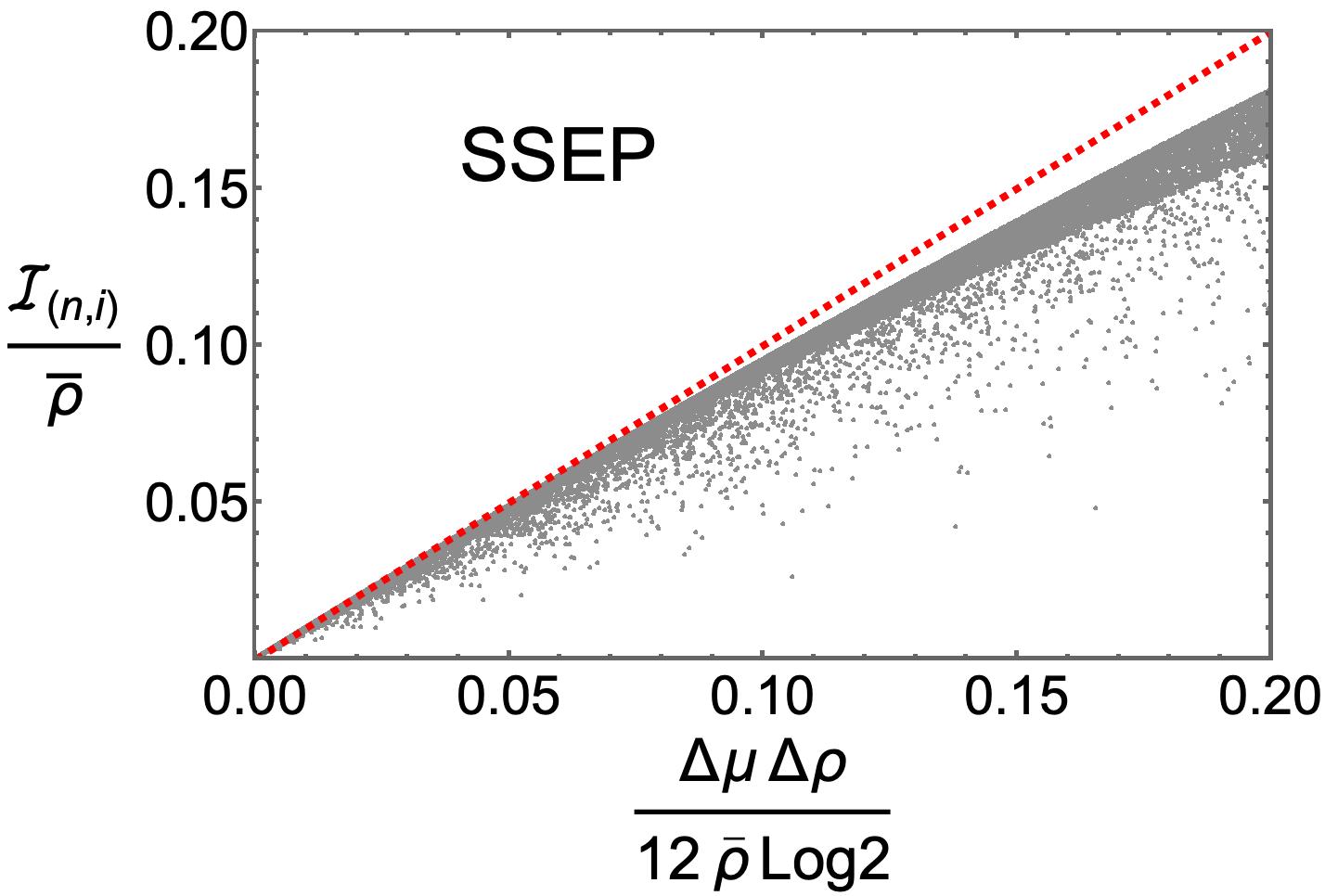}}
  \subfigure{\includegraphics[scale=0.3]{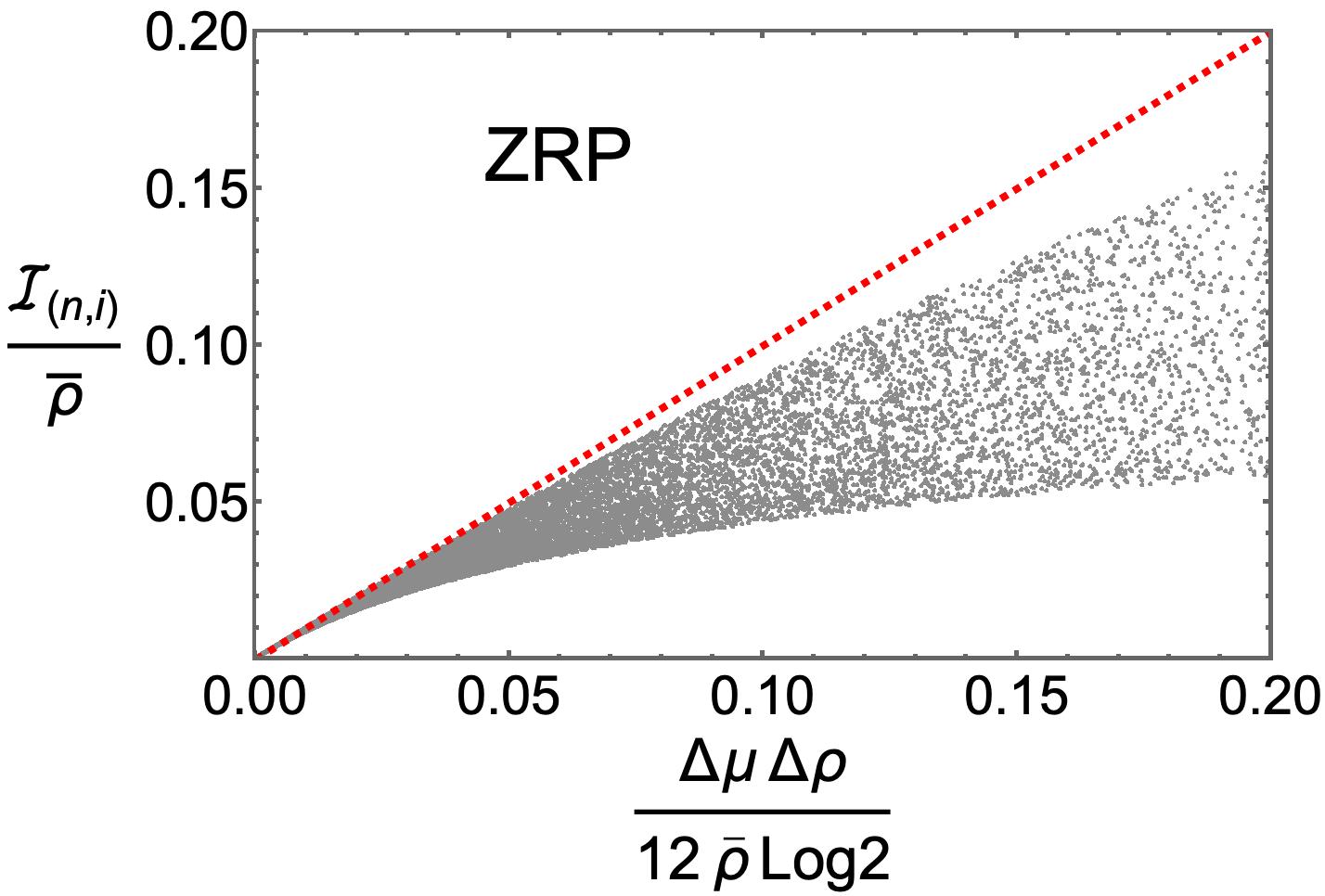}}
  \subfigure{\includegraphics[scale=0.3]{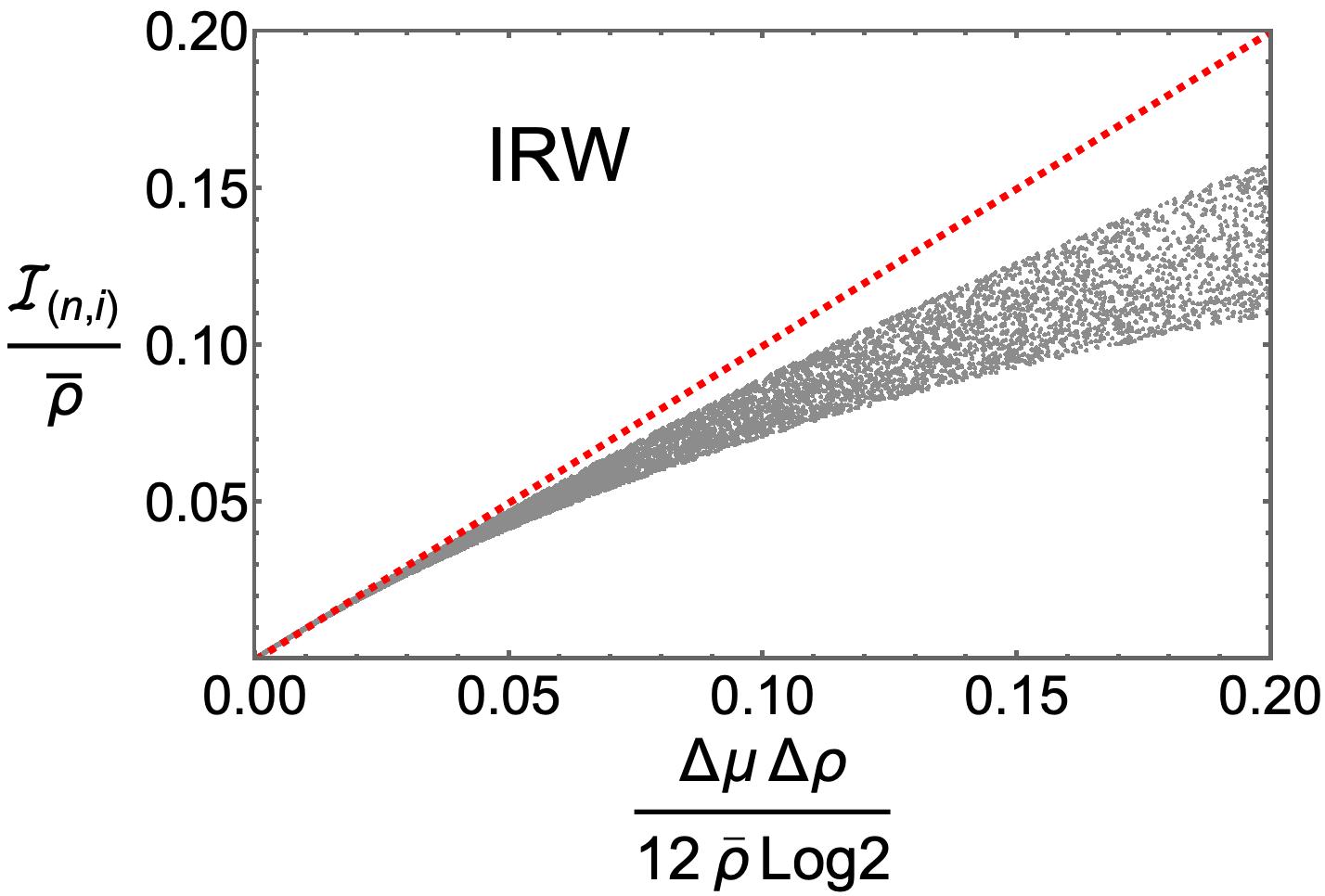}}
  \subfigure{\includegraphics[scale=0.3]{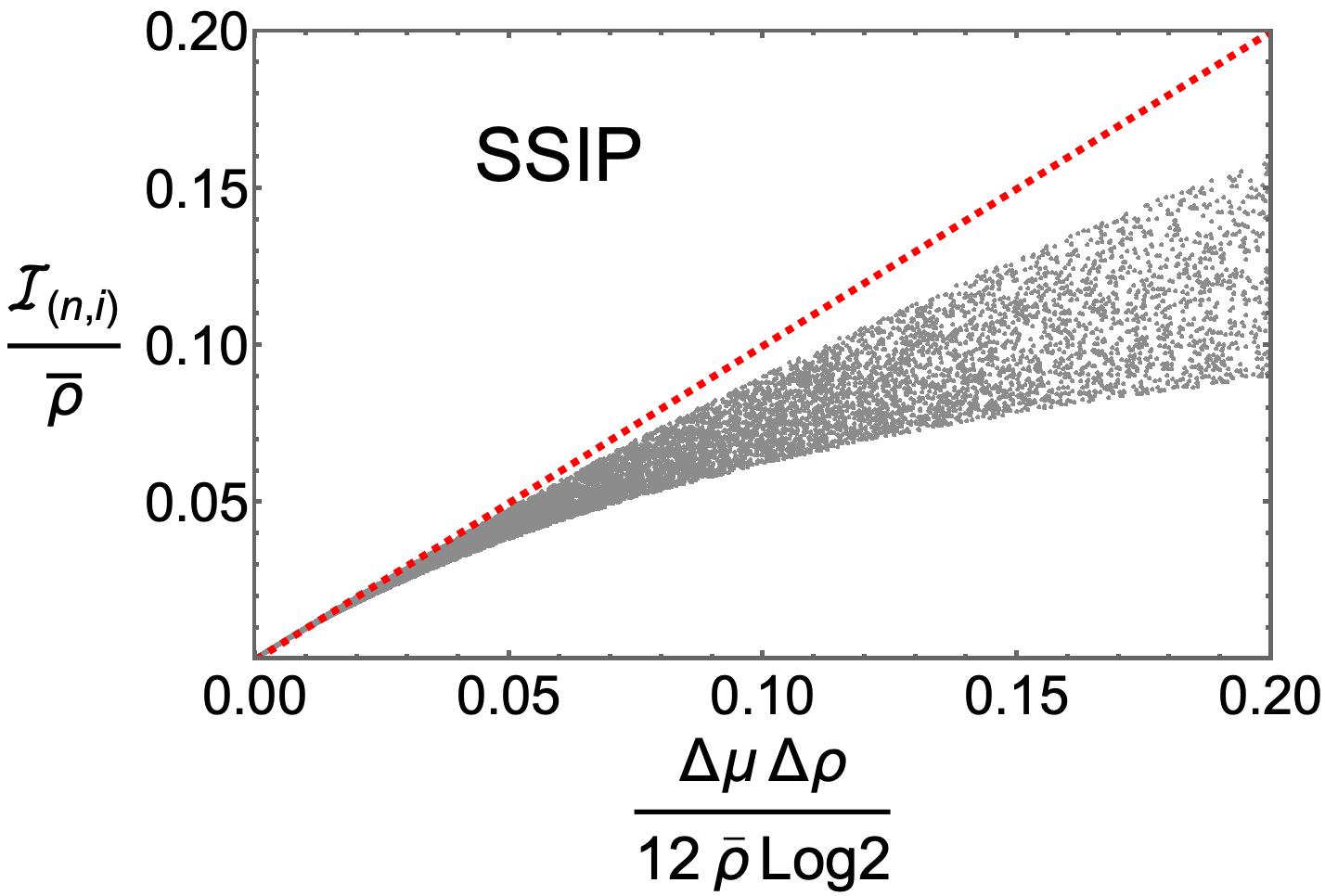}}
\centering
\caption{\ps{In this plot, we have rescaled both sides of \eqref{appen-ineqq-eq-1} by $\bar{\rho}  $ and plotted $\mathcal{I} _{\left( n, i\right) } \left( \bar{\rho}, \Delta \rho \right) \Big/ \bar{\rho} $ vs $\Delta \mu \Delta \rho \Big/ 12 \bar{\rho} \ln 2$. As discussed below Eq.~\eqref{nnea-eq-1n1}, this rescaling ensures that the near-equilibrium relation \eqref{nnea-eq-1n1} is more clearly visible. The red line corresponds to the upper bound in \eqref{appen-ineqq-eq-1} for the flat prior case. We clearly see that the bound becomes an equality when close to equilibrium.}}
\label{fig-NN-new}
\end{figure} 
Hence for SSEP, we have rigorously derived that the bound \eqref{appen-ineqq-eq-1} is valid. \ps{Moreover, this upper bound turns into an equality in the near-equilibrium regime. To see this, we observe that for $\Delta \rho \ll \bar{\rho}$ and $\Delta \rho \ll (1-\bar{\rho})$, one can expand $\mathcal{I} _{\left( n, i\right) } \left( \bar{\rho}, \Delta \rho \right)$ in Eq.~\eqref{I-int-eqn} to obtain
\begin{align}
\mathcal{I} _{\left( n, i\right) } \left( \bar{\rho}, \Delta \rho \right) \simeq \frac{\Delta \rho^2 }{6 ~ \bar{\rho}(1-\bar{\rho})\ln 2 } + \mathcal{O}(\Delta \rho ^4).
\end{align}
This matches with our general near-equilibrium expression in Eq.~\eqref{supp-general-bd-eq-5} with $\sigma _2 (\bar{\rho}) = \bar{\rho}(1-\bar{\rho})$. Using this with Eq.~\eqref{safiqp}, we obtain
\begin{align}
\mathcal{I} _{\left( n, i\right) } \left( \bar{\rho}, \Delta \rho \right) \approx \frac{\Delta \mu~ \Delta \rho}{12 \ln 2}. \label{nnea-eq-1n1}
\end{align}
This relation is valid only when $\Delta \rho \ll \bar{\rho}$ and $\Delta \rho \ll (1-\bar{\rho})$. Hence for small $\Delta \rho$, the upper bound \eqref{mqobxt} becomes an equality. We have illustrated this in Figure~\ref{fig-NN-new} where we have plotted the two sides of \eqref{mqobxt} along the two axes. Note that $\Delta \mu$ will also be small when $\bar{\rho} \gtrsim \Delta \rho $ and $\Delta \rho \to 0^+$. However, the relation in Eq.~\eqref{nnea-eq-1n1} will no longer hold under these conditions. Therefore, for a systematic comparison, we need to normalise both sides of Eq.~\eqref{nnea-eq-1n1} by $\bar{\rho}$ and plot $\mathcal{I} _{\left( n, i\right) } \left( \bar{\rho}, \Delta \rho \right) \Big/ \bar{\rho} $ vs $\Delta \mu \Delta \rho \Big/ 12 \bar{\rho} \ln 2$. This rescaling ensures that Eq.~\eqref{nnea-eq-1n1} will be satisfied under correct conditions.}

\ps{Furthermore to make this plot, we generate a random pair of $(\bar{\rho}, \Delta \rho)$ values by selecting $\bar{\rho}$ uniformly from the range $[0,1]$, and then choosing $\Delta \rho$ uniformly from $[0, \bar{\rho}]$. Using this pair, we then calculate the positional information and $\Delta \mu~\Delta \rho$ and obtain one single data point (gray point) in top left in Figure~\ref{fig-NN-new}. This is then repeated $10^4$ times to get the complete plot.} The red dashed line indicates our upper bound. We clearly see that the positional information remains below this red line across all parameter values.

%We have shown this in Figure~\ref{fig-NN-new}. Note that $\Delta \mu$ will also be small when $\bar{\rho} \gtrsim \Delta \rho $ and $\Delta \rho \to 0^+$. However, the relation in Eq.~\eqref{nnea-eq-1n1} will no longer hold under these conditions. Therefore, for a systematic comparison, we need to normalise both sides of Eq.~\eqref{nnea-eq-1n1} by $\bar{\rho}\Delta \rho$ and plot $\frac{\mathcal{I} _{\left( n, i\right) } \left( \bar{\rho}, \Delta \rho \right)}{\bar{\rho} \Delta \rho}~\text{vs}~\frac{\Delta \mu~ \Delta \rho}{12\bar{\rho} \Delta \rho \ln 2}$. This ensures that Eq.~\eqref{nnea-eq-1n1} will be satisfied under correct conditions.
\begin{figure}[t]
	\includegraphics[scale=0.6]{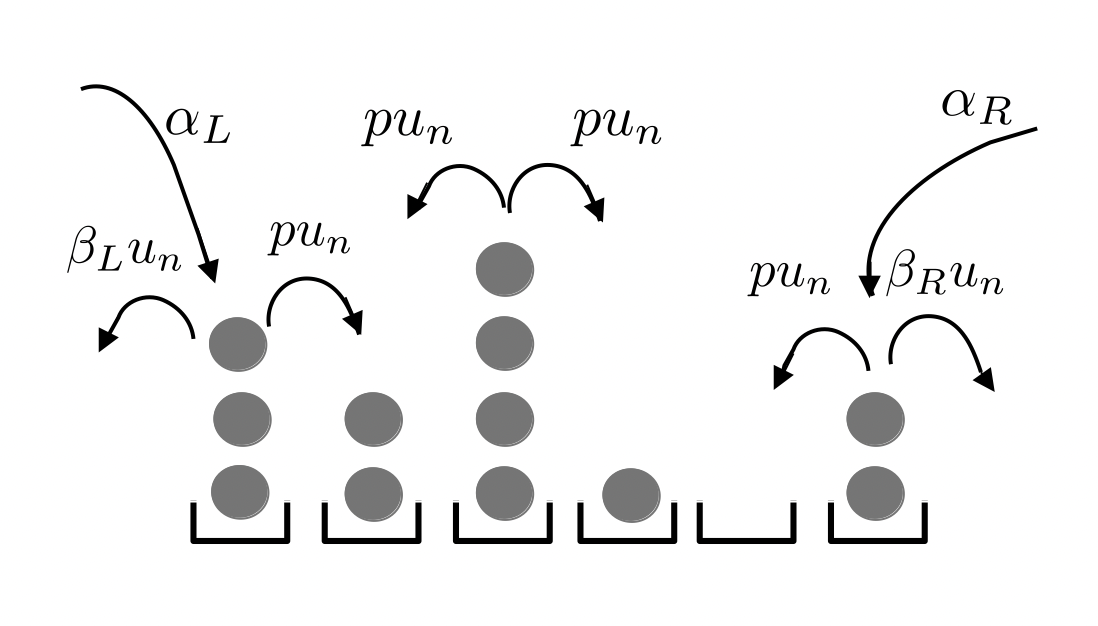}
	\centering
	\caption{Zero-range process with open boundaries} 
	\label{ZRP-fig}
\end{figure}

\section{Positional information for the boundary driven ZRP}
We now demonstrate positional information for another model, namely the open zero-range process (ZRP) and test our upper bound. The model consists of $N$ lattice sites and each site can accommodate an arbitrary non-negative number of particles. This is different than the SSEP where each site can have at most a single particle. From any bulk site $i$, a particle can jump to  either of its neighbouring sites with a rate $p u_{n_i}$, where $u_{n_i}$ is a non-negative function of the number of particles $n_i$. At the boundaries $(i=1 \text{ and }i=N)$, the dynamics are modified to allow for the addition (removal) of a particle with rate $\alpha _L~(\beta _L u_{n_1})$ for the left boundary and with rate $\alpha _R~(\beta _R u_{n_N})$ for the right one (see Figure \ref{ZRP-fig}). Following \cite{Levine2005}, whenever steady state exists, the probability to find $n_i=n$ number of particles in the $i$-th site is given by
\begin{align}
P(n\big|i)= \mathcal{N}_i (z_i)^n \prod _{n_k=1}^{n}\frac{1}{u_{n_k}},~~~~\text{with }z_i = \frac{\alpha _L}{\beta _L}-\left( \frac{\alpha _L}{\beta _L}-\frac{\alpha _R}{\beta _R} \right) \left( \frac{i}{N} \right), \label{eq-ZRP-1}
\end{align}
where $\mathcal{N}_k$ is the normalisation factor. For the case of $u_n=1$, the probability $P(n\big|i)$ takes the form
\begin{align}
P(n \big|i) = (z_i)^n(1-z_i), ~~~~\text{with }z_i = \frac{\alpha _L}{\beta _L}-\left( \frac{\alpha _L}{\beta _L}-\frac{\alpha _R}{\beta _R} \right) \left( \frac{i}{N} \right). \label{eq-ZRP-2}
\end{align}
As observed in the case of the SSEP, managing summations over $i$ can be challenging. To simplify this, it is helpful to introduce the variable change $x=i/N$ for large $N$ and continue the analysis in terms of $x$. Rewriting Eq.~\eqref{eq-ZRP-2}
\begin{align}
P(n \big| x) \simeq \left[ z(x)\right]^n \left[ 1-z(x)\right], ~~~~\text{with }z(x) = \frac{\alpha _L}{\beta _L}-\left( \frac{\alpha _L}{\beta _L}-\frac{\alpha _R}{\beta _R} \right) x. \label{eq-ZRP-1ka3}
\end{align}
Since, we are interested in calculating the positional information in terms of the variables $\bar{\rho} = (\rho _L +\rho _R)/2$ and $\Delta \rho =(\rho _L -\rho _R)/2$, where $\rho _L =\langle n(0) \rangle$ and $\rho _R =\langle n(1) \rangle$
are average densities at the two ends
\begin{align}
\rho _L = \frac{\alpha _L}{\beta  _L-\alpha _L},~~~\rho _R = \frac{\alpha _R}{\beta  _R-\alpha _R}, \label{eq-ZRP-4}
\end{align}
we rewrite Eq.~\eqref{eq-ZRP-1ka3} in the following manner
\begin{align}
P(n \big| x) \simeq \left[ z(x)\right]^n \left[ 1-z(x)\right],~~~\text{with }z(x) = c_+ -x \left( c_+-c_- \right). \label{eq-ZRP-3}
\end{align}
where $c_{\pm} = (\bar{\rho} \pm \Delta \rho)/(1+\bar{\rho} \pm \Delta \rho)$. For this model, the average density in the bulk takes a non-linear form 
\begin{align}
\rho(x) = \frac{\bar{\rho} (1+\bar{\rho})-\Delta \rho~ (2x-1+\Delta \rho)}{(1+\bar{\rho})+\Delta \rho~ (2x-1)}. \label{qahyq}
\end{align}
In the remaining part of this section, we will demonstrate that even with this non-linear profile, the bound \eqref{appen-ineqq-eq-1} remains still valid. To see this, we use the probability in Eq.~\eqref{eq-ZRP-3} and calculate the entropy
\begin{align}
S[P(n \big| x)] &= -\sum _{n=0}^{\infty} P(n \big| x) ~\log _2 P(n \big| x) = -\frac{z(x) \log _2 z(x)+ \left(  1-z(x) \right) \log _2 \left(  1-z(x) \right)}{\left(  1-z(x) \right)}. \label{eq-ZRP-6}
\end{align}
Taking average with respect to the prior $P_x(x)=1$ gives
\begin{align}
& \langle  S[P(n \big| x)]  \rangle_x \nonumber \\
&= -\int _{0}^{1}dx~ \frac{z(x) \log _2 z(x)+ \left(  1-z(x) \right) \log _2 \left(  1-z(x) \right)}{\left(  1-z(x) \right)}, \nonumber \\
& =- \frac{1}{\ln 2 \left(c_+-c_- \right)} \Big[ \text{Li}_2(1-c_+) -\text{Li}_2(1-c_-)  -c_+ \ln c_+  + c_- \ln c_--(1-c_+) \ln (1-c_+)+(1-c_-) \ln (1-c_-)  \Big],
\label{eq-ZRP-7}
\end{align}
where $\text{Li}_2(y)$ denotes the poly-logarithmic function. Next we calculate the other entropy term $S[P_n(n)]$ in Eq.~\eqref{appen-avg-information} for which we need the following probability 
\begin{align}
P_n(n) &= \int _{0}^{1} dx ~P(n\big|x) P_x(x)= \frac{f(n)}{\left(c_+-c_- \right)},~~~\text{with }f(n) = \frac{c_+^{n+1}-c_-^{n+1}}{n+1}- \frac{c_+^{n+2}-c_-^{n+2}}{n+2}.
\label{eq-ZRP-5}
\end{align}
This yields
\begin{align}
S[P_{n}(n)] = -\sum _{n=0}^{\infty} P_{n}(n) \log _2 P_{n}(n) = \log _2 \left( c_+-c_-\right) - \frac{1}{ \left( c_+-c_-\right) } \sum _{n=0}^{\infty} f(n) \log _2 f(n). \label{bb-fank781r}
\end{align}
Using Eqs.~\eqref{eq-ZRP-7} and \eqref{bb-fank781r}, the expression of the positional information turns out to be
\begin{equation}
\scalebox{1}{$
\begin{split}
\mathcal{I} _{\left( n, i\right) } \left( \bar{\rho}, \Delta \rho \right) & = \frac{1}{\ln 2~(c_+-c_-)} \Big[ \text{Li}_2(1-c_+) -\text{Li}_2(1-c_-)  -c_+ \ln c_+  + c_- \ln c_--(1-c_+) \ln (1-c_+) \Big. \\
& ~~~~~~~~~~\Big.
+(1-c_-) \ln (1-c_-)  -\sum _{n=0}^{\infty}f(n) \ln f(n) \Big] + \log _2(c_+-c_-).
\end{split}$} \label{bb-fank781r2ed}
\end{equation}
We have compared this result with the numerical simulations in Figure~\ref{SM-avginf-fig}, and found a good agreement between them. Unlike in SSEP, the positional information for ZRP decreases monotonically with $\bar{\rho}$ and vanishes for large $\bar{\rho}$. For larger values of $\bar{\rho}$ but with fixed $\Delta \rho$, both boundary sites have a large number of particles available for hopping in the bulk, as there is no exclusion. Hence, in the steady-state, we expect the same number of particles in the bulk as well as in the boundaries. This can also be seen from Eq.~\eqref{qahyq} where the density becomes independent of $x$ for large enough $\bar{\rho}$. Therefore, the probability $P(x|n)$ is broad with respect to $x$, and consequently the entropy $S[P(x|n)]$ takes a large value. From Eq.~\eqref{appen-avg-information-new}, this would mean that the positional information is small. In fact our study shows that, in processes with no exclusion, $\mathcal{I} _{\left( n, i \right) } \left( \bar{\rho} , \Delta \rho \right)$ exhibits a monotonic decay.

Our expression in Eq.~\eqref{bb-fank781r2ed} is also consistent with the one derived with perturbative approach in Eq.~\eqref{appen-exp-2terms}. This can be verified by expanding Eq.~\eqref{bb-fank781r2ed} in $\Delta \rho $
\begin{equation}
\begin{split}
\mathcal{I} _{\left( n, i\right) } \left( \bar{\rho}, \Delta \rho \right)  \approx \frac{\Delta \rho ^2 }{6 \ln 2~ \bar{\rho} (1+ \bar{\rho} ) }+ \frac{\left[ 3+\bar{\rho }(11+7 \bar{\rho} ) \right] \Delta \rho ^4}{180 \ln 2 ~\bar{\rho }^3(1+\bar{\rho })^3}.
\end{split}
\label{supp-ZRP-paper-eq-2}
\end{equation}
One essentially gets the same expression also from Eq.~\eqref{appen-exp-2terms} by plugging the central moments from Eq.~\eqref{eq-ZRP-3}.  

Now that we have obtained an exact expression for the positional information, we will employ it to test the bound in Eq.~\eqref{appen-ineqq-eq-1}. First, we compute the chemical potential difference using Eq.~\eqref{kmayq} for which we need the first derivative of the free energy. In this regard, we follow a procedure similar to that of the SSEP and employ the fluctuation-response relation in Eq.~\eqref{FTT-appen}. 
\begin{align}
a_2(\bar{\rho}) = \frac{1}{\sigma _2(\bar{\rho})} = \frac{1}{\bar{\rho}(1+\bar{\rho})}, \implies a_1(\bar{\rho}) = \ln \left( \frac{\bar{\rho}}{1+\bar{\rho}} \right).
\end{align}
The resulting expression for $\Delta \mu$ is now found to be
\begin{align}
\Delta \mu = \ln \left( \frac{\bar{\rho}+ \Delta \rho}{1+\bar{\rho}+\Delta \rho }  \right)-\ln \left( \frac{\bar{\rho}- \Delta \rho}{1+\bar{\rho}-\Delta \rho }  \right). \label{xahvuyqZRP}
\end{align}  
We now utilize these exact expressions to plot $\mathcal{I} _{\left( n, i \right) } \left( \bar{\rho} , \Delta \rho \right) $ and $  (\Delta \mu ~\Delta \rho )/( 12 \ln 2)$ for all possible values of $\bar{\rho}$ and $\Delta \rho$ in Figure~\ref{fig-NN-new}. \ps{Here again, these quantities are rescaled by $\bar{\rho} $ to retrieve their equality in the appropriate parameter regime. Moreover, the plot is generated in the following manner. We choose a random pair of $(\bar{\rho}, \Delta \rho)$ values by selecting $\bar{\rho}$ uniformly from the range $[0,5]$, and then choosing $\Delta \rho$ uniformly from $[0, \bar{\rho}]$. With this pair, we then calculate the positional information and $\Delta \mu~\Delta \rho$ and obtain one single data point (gray point) in top right in Figure~\ref{fig-NN-new}. We next repeat this $10^4$ times to get the full plot.} Across all these parameter values, we see that the bound \eqref{appen-ineqq-eq-1} is satisfied, and it is saturated in the near-equilibrium limit. To sum up, this section showcases an example of a model with non-linear density profile, where \eqref{appen-ineqq-eq-1} remains valid across all parameter values.

\begin{table*}
\centering
\begin{tabular*}{\textwidth}{@{\extracolsep{\fill}} lcc}
\hline 
\addlinespace
\textbf{Model} & \bm{$\Delta \mu$} & \bm{$\mathcal{I} _{\left( n, i\right) } \left( \bar{\rho}, \Delta \rho \right) $} \\
\addlinespace
\hline 
\addlinespace
SSEP & $\ln \left( \frac{\bar{\rho}+ \Delta \rho}{1-\bar{\rho}-\Delta \rho }  \right)-\ln \left( \frac{\bar{\rho}- \Delta \rho}{1-\bar{\rho}+\Delta \rho }  \right)$ & $\begin{aligned}
& \frac{1}{\ln 16 ~\Delta \rho} \Big[ \mathcal{Y}(\bar{\rho}, \Delta \rho) +\mathcal{Y}(1-\bar{\rho}, \Delta \rho) -2 \Delta \rho  \Big]  - \left( 1-\bar{\rho} \right) \log _2  \left( 1-\bar{\rho} \right) -\bar{\rho }\log _2 \bar{\rho }
\end{aligned}$ \\
[7pt]
\addlinespace
\hline
\addlinespace
ZRP & $ \ln \left( \frac{\bar{\rho}+ \Delta \rho}{1+\bar{\rho}+\Delta \rho }  \right)-\ln \left( \frac{\bar{\rho}- \Delta \rho}{1+\bar{\rho}-\Delta \rho }  \right)$ &  $\begin{aligned} & \frac{1}{\ln 2(c_+-c_-)} \Big[ \text{Li}_2(1-c_+) -\text{Li}_2(1-c_-) + c_- \ln c_-   -(1-c_+) \ln (1-c_+) \Big.  \\
  & +(1-c_-) \ln (1-c_-)   \Big.- c_+ \ln c_+ -\sum _{n=0}^{\infty}f(n) \ln f(n) \Big] + \log _2(c_+-c_-)
\end{aligned}$~~~~
 \\
 \addlinespace
[10pt]
\hline
\addlinespace
IRW &  $\ln(\bar{\rho}+\Delta \rho)-\ln(\bar{\rho}-\Delta \rho)~~~~~~$ & $\begin{aligned}
 \frac{1}{\ln 16 ~\Delta \rho}\Big[ \mathcal{Y}(\bar{\rho}, \Delta \rho)  - 6 \bar{\rho} ~\Delta \rho  \Big]  -\sum _{n=0}^{\infty}\frac{g(n)}{n!} \log _2 g(n)
\end{aligned} $ ~~~~~~~~~~~~~~~~~~~~~~~~~~~~~~~~~~ \\ 
 \addlinespace
[7pt]
\hline
 \addlinespace
SSIP & $ \ln \left( \frac{\bar{\rho}+ \Delta \rho}{m+\bar{\rho}+\Delta \rho }  \right)-\ln \left( \frac{\bar{\rho}- \Delta \rho}{m+\bar{\rho}-\Delta \rho }  \right)$ & $\begin{aligned}
& \frac{1}{\ln 16 ~\Delta \rho}\Big[ \mathcal{Y}(\bar{\rho}, \Delta \rho) - \mathcal{Y}(m+\bar{\rho}, \Delta \rho)  + 2m ~\Delta \rho  \Big] +(m-1) \log _2 m\\
&-\sum _{n=0}^{\infty}\frac{m \Gamma (m+n)}{\Gamma(n+1) \Gamma(m)}\mathcal{S}(n) \log _2 \mathcal{S}(n)
\end{aligned}$~~~~~~~~~~~~~~~~
 \\
 \addlinespace
\hline
\end{tabular*}
\caption{Summary of the expressions of the positional information for the flat prior and chemical potential difference for different boundary-driven systems studied in the letter.} 
%Furthermore, the forms of different functions used to represent them are provided in \bluew{\cite{expp}}.
\label{Table-chempot}
\end{table*}
\subsection{Independent random walkers}
\noindent
In our preceding analysis, we considered a special case of zero-range process where the rate $u_n=1$ was chosen to be constant. We now provide another example of $u_n = n$ where an exact expression for the positional information can be obtained and the bound can be assessed. This example corresponds to the case of independent random walkers (IRW) \cite{Levine2005}. We use $u_n=n$ in Eq.~\eqref{eq-ZRP-1} and obtain the conditional probability as
\begin{align}
P(n|x) = \frac{\left[ \rho(x)\right]^n}{n!}\exp\left[- \rho (x)\right],~~~\text{where }\rho(x) = \bar{\rho} -(2x-1)\Delta \rho, \label{eq-ZRP-13}
\end{align}
where we have written the expression in terms of the $x$ variable. The marginal probability $P_n(n)$ follows to be 

\begin{align}
P_n(n) =  \frac{g(n)}{n!},~~~~\text{with }g(n) =
\frac{\Gamma \left( n+1, \bar{\rho}-\Delta \rho  \right)-\Gamma \left( n+1, \bar{\rho}+\Delta \rho  \right)}{2 ~\Delta \rho}.\label{eq-ZRP-15}
\end{align}
We now have both probabilities required in the definition of positional information in Eq.~\eqref{appen-avg-information}.
One can now proceed in the same manner as before and calculate the two entropies as
\begin{align}
& S[P_n(n)]  = \sum _{n=0}^{\infty} P_n(n)\log _2 n!-\sum _{n=0}^{\infty}\frac{g(n)}{n!} \log _2 g(n), \\
& \langle  S[P(n \big| x)]  \rangle_x  =  \frac{1}{\ln 16 ~\Delta \rho} \Big[ (\bar{\rho}-\Delta \rho)^2 \ln (\bar{\rho}-\Delta \rho)- (\bar{\rho}+\Delta \rho)^2 \ln (\bar{\rho}+\Delta \rho)  + 6 \bar{\rho} ~\Delta \rho \Big] + \sum _{n=0}^{\infty} P_n(n)\log _2 n!
\end{align}
from which the expression of positional information can be calculated to be
%\begin{align}
%\mathcal{I}_{(n,x)}(\bar{\rho}, \Delta \rho) = \frac{1}{\ln 16 ~\Delta \rho} \Big[ (\bar{\rho}+\Delta \rho)^2 \ln (\bar{\rho}+\Delta \rho)- (\bar{\rho}-\Delta \rho)^2 \ln (\bar{\rho}-\Delta \rho)  - 6 \bar{\rho} ~\Delta \rho \Big] -\sum _{n=0}^{\infty}\frac{g(n)}{n!} \log _2 g(n),
%\end{align}
\begin{align}
\mathcal{I}_{(n,i)}(\bar{\rho}, \Delta \rho) = \frac{1}{\ln 16 ~\Delta \rho} \Big[ \mathcal{Y}(\bar{\rho}, \Delta \rho)  - 6 \bar{\rho} ~\Delta \rho \Big] -\sum _{n=0}^{\infty}\frac{g(n)}{n!} \log _2 g(n),
\end{align}
with functions $\mathcal{Y}(\bar{\rho}, \Delta \rho)$ and $g(n)$ defined respectively in Eqs.~\eqref{y-fun-eq} and \eqref{eq-ZRP-15}. Similarly, the chemical potential difference driving this system is
\begin{align}
\Delta \mu = \ln(\bar{\rho}+\Delta \rho)-\ln(\bar{\rho}-\Delta \rho) \label{xahvuyqIRW}
\end{align}
We now have at our disposal all the quantities needed to test the upper bound \eqref{appen-ineqq-eq-1} on the positional information. Plotting $\mathcal{I} _{\left( n, i\right) } \left( \bar{\rho}, \Delta \rho \right) \Big/ \bar{\rho} \Delta \rho$ vs $\Delta \mu \Big/ 12 \bar{\rho} \ln 2$ in Figure~\ref{fig-NN-new}, we have illustrated that the bound is satisfied even for this model. This model presents a third solvable example where the upper bound is valid. \ps{The data points for this model were generated using the same procedure as for the ZRP.} 

\begin{figure}[t]
	\includegraphics[scale=0.34]{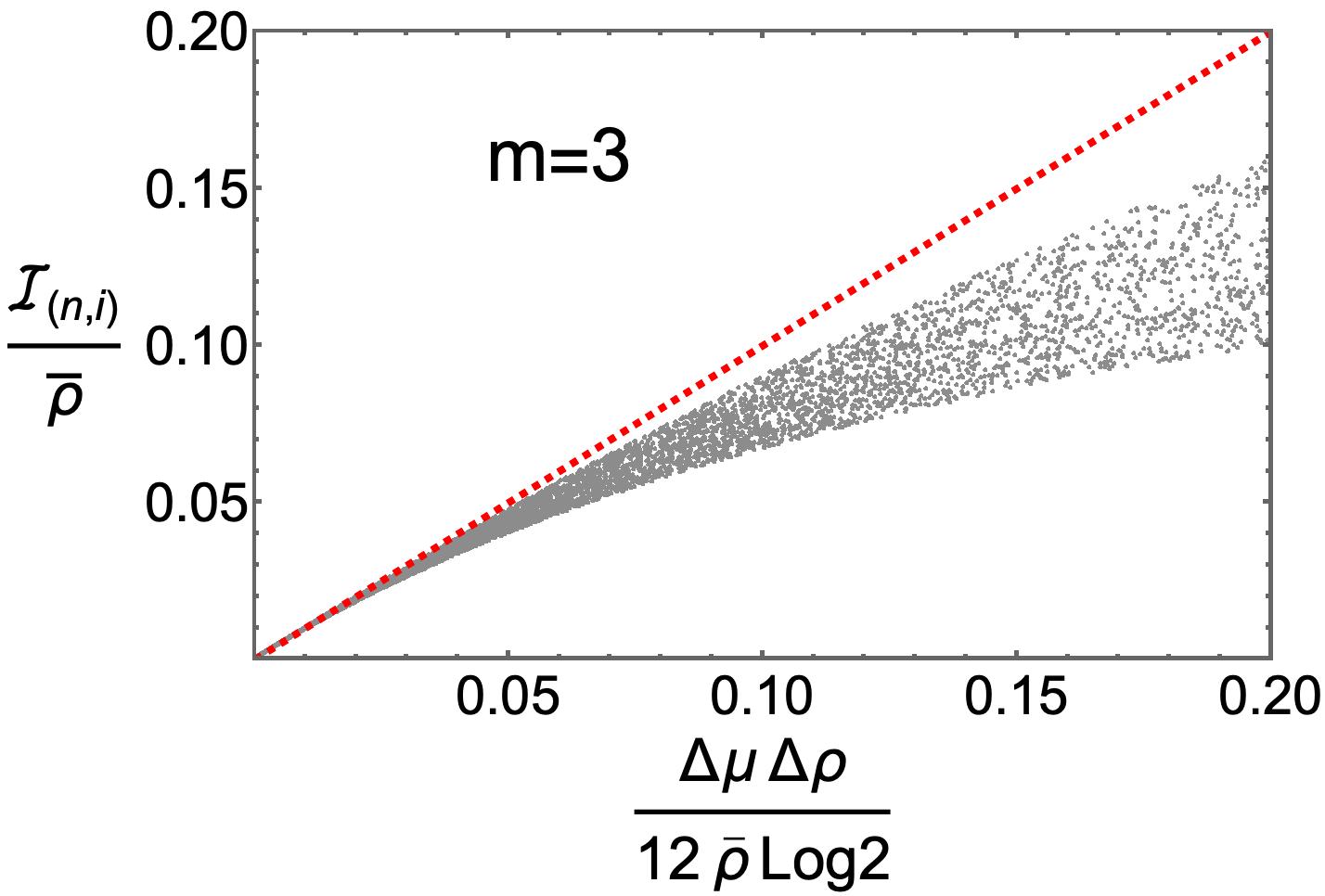}
	\includegraphics[scale=0.34]{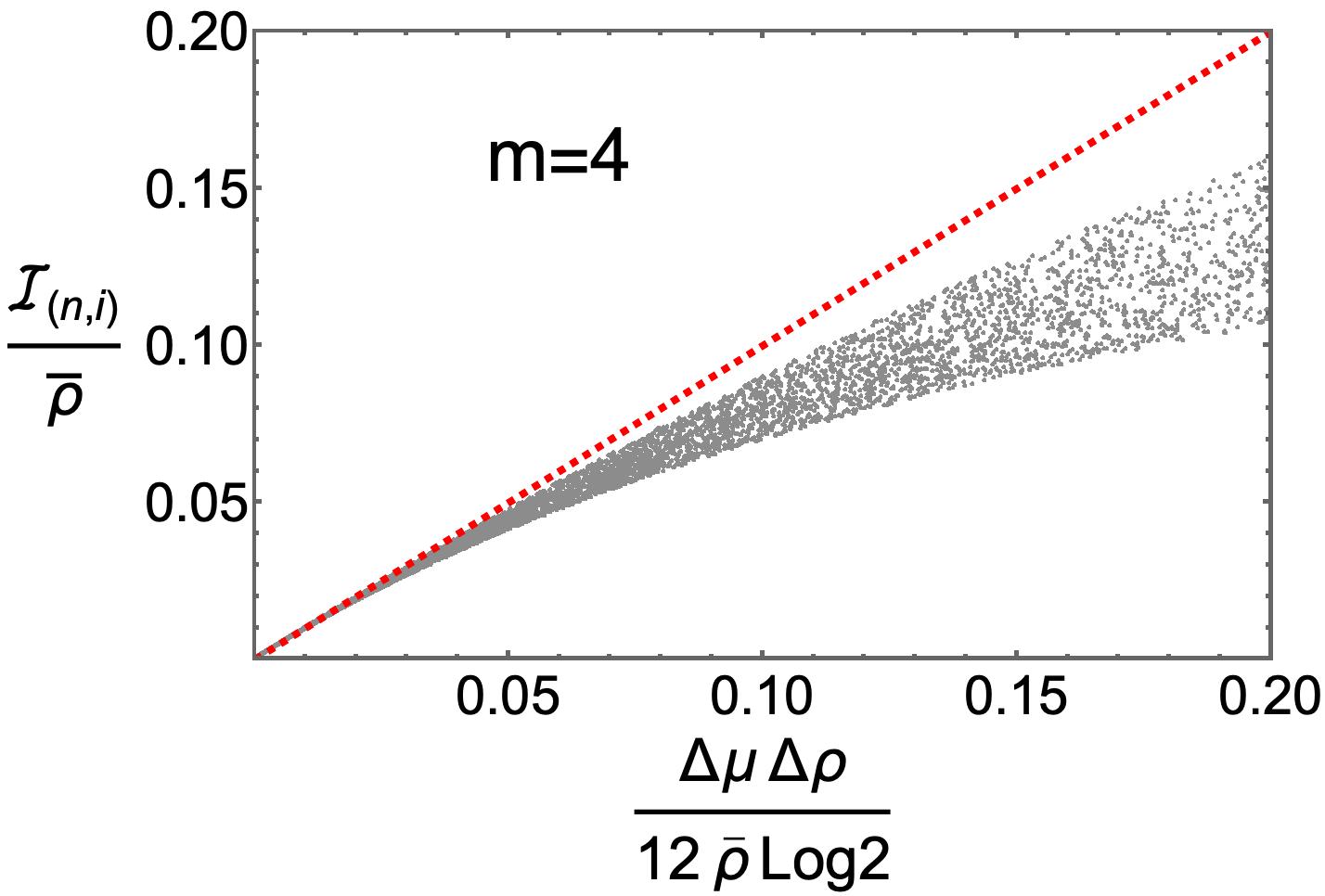}
	\centering
	\caption{Illustration of the upper bound on the positional information for SSIP model for two different values of the parameter $m$. The corresponding expressions of $\mathcal{I} _{\left( n, i\right) } \left( \bar{\rho}, \Delta \rho \right)$ and $\Delta \mu$ are given in Eqs.~\eqref{appen-SSIP-eq-6} and \eqref{appen-SSIP-eq-7} respectively.} 
	\label{SSIP-other-fig}
\end{figure}
\section{Positional information for the boundary-driven SSIP}
\redw{In all models examined thus far, the jump rate was either independent of the particle numbers or solely dependent on the occupation number of the source site from which the jump takes place. In this section, we will consider the SSIP model where the jump rate depends on the occupation numbers of both the source and the target sites \cite{PhysRevE.90.052143, 10.3150/21-BEJ1390}. Like in ZRP, each lattice site is capable of accommodating an arbitrary non-negative number of particles. In the bulk site, a particle can jump from $i \to j$ [with $j=(i \pm 1)$] with a rate $p n_i(n_j+m)$ where $m~(>0)$ is a parameter in the model. At the boundaries $(i=1 \text{ and }i=N)$, a particle  can be added with rate $\alpha _L (n_1+m)$ for the left boundary and $\alpha _R~(n_N+m)$ for the right one. Similarly, a particle, if present, can be removed from these sites with rates
$\beta _L n_1$ and $\beta _R n_N$ respectively. The conditional probability to observe $n$ particles at location $i = xN$ is \cite{PhysRevE.90.052143}
\begin{align}
P(n|x) =\left[ \frac{m^m \Gamma (m+n)}{n!~ \Gamma(m)} \right]~\frac{\left[ \rho(x) \right] ^n}{\left[ m+\rho(x)  \right]^{m+n}} ,~~~\text{where }\rho(x) = \bar{\rho} -(2x-1)\Delta \rho, \label{appen-SSIP-eq-1}
\end{align}
from which the marginal probability $P_n(n)$ follows to be
\begin{align}
& P_n(n) =  \frac{m ~\Gamma (m+n)~\mathcal{S}(n)}{n!~ \Gamma(m)} ,~~~\text{where } \label{appen-SSIP-eq-2}  \\
& \mathcal{S}(n) =  \frac{1}{2(n+1)\Delta \rho}  \left[ \left( \frac{\bar{\rho}+\Delta \rho}{m}\right) ^{n+1} {}_2F_1 \left( 1+n,m+n; 2+n;  -\frac{\bar{\rho}+\Delta \rho}{m}\right)  -\left( \frac{\bar{\rho}-\Delta \rho }{m}\right) ^{n+1}  \right.\nonumber \\
& ~~~~~~~~~~~~~~~~~~~~~~~~~~\left. \times~ {}_2F_1 \left( 1+n,m+n; 2+n;  -\frac{\bar{\rho}+\Delta \rho}{m}\right) \right]. \label{appen-SSIP-eq-3}
\end{align}
Here ${}_2F_1 \left( 1+n,m+n; 2+n; z \right) $ stands for the hypergeometric function. Now that we possess both probabilities, we can compute the two entropies to be
\begin{align}
S[P_n(n)] & = -\left[  \log _2 m +\sum _{n=0}^{\infty} P_n(n) \log _2 \left( \frac{\Gamma(m+n)~\mathcal{S}(n)}{n!~\Gamma(m)}  \right)  \right],  \label{appen-SSIP-eq-4} \\
\langle S[P(n|x)] \rangle _x &  = \frac{1}{\ln 16 ~\Delta \rho} \Big[  \mathcal{Y}(m+\bar{\rho}, \Delta \rho)  -\mathcal{Y}(\bar{\rho}, \Delta \rho) -2 m \Delta \rho \Big]-\sum _{n=0}^{\infty} P_n(n) \log _2 \left( \frac{\Gamma(m+n)~m^m}{n!~\Gamma(m)}  \right).  \label{appen-SSIP-eq-5}
\end{align}
Using these two expressions, the positional information can be calculated to be
\begin{align}
\mathcal{I} _{\left( n, i\right) } \left( \bar{\rho}, \Delta \rho \right) = & \frac{1}{\ln 16 ~\Delta \rho}\Big[ \mathcal{Y}(\bar{\rho}, \Delta \rho) - \mathcal{Y}(m+\bar{\rho}, \Delta \rho)  + 2m ~\Delta \rho  \Big] +(m-1) \log _2 m -\sum _{n=0}^{\infty}\frac{m~ \Gamma (m+n)~\mathcal{S}(n)}{\Gamma(n+1) \Gamma(m)} \log _2 \mathcal{S}(n),  \label{appen-SSIP-eq-6}
\end{align}
with functions $\mathcal{Y}(\bar{\rho}, \Delta \rho)$ and $\mathcal{S}(n)$ given respectively in Eq.~\eqref{y-fun-eq} and \eqref{appen-SSIP-eq-3}. Similarly, the chemical potential difference driving the system is
\begin{align}
\Delta \mu = \ln \left( \frac{\bar{\rho}+\Delta \rho}{m+\bar{\rho}+\Delta \rho}   \right) - \ln \left( \frac{\bar{\rho}-\Delta \rho}{m+\bar{\rho}-\Delta \rho}   \right).  \label{appen-SSIP-eq-7}
\end{align}
We are now equipped with all the necessary elements to assess the upper bound \eqref{appen-ineqq-eq-1} on positional information. In Figure \ref{fig-NN-new}, we have again plotted $\mathcal{I} _{\left( n, i\right) } \left( \bar{\rho}, \Delta \rho \right) \Big/ \bar{\rho} \Delta \rho$ vs $\Delta \mu \Big/ 12 \bar{\rho} \ln 2$ for $m=2$ for different values of $\Delta \rho $ and $\bar{\rho}$. \ps{The data points for this model were generated using the same procedure as for the ZRP.} Across all these values, we once again find that our bound is satisfied. We have also verified this for other values of $m$ in Figure \ref{SSIP-other-fig}. This consistency further reinforces the conjecture that our bound  holds for general boundary-driven systems.}
\section{Non-flat prior}
\label{non-flat}
\begin{figure}[t]
  \centering
  \subfigure{\includegraphics[scale=0.34]{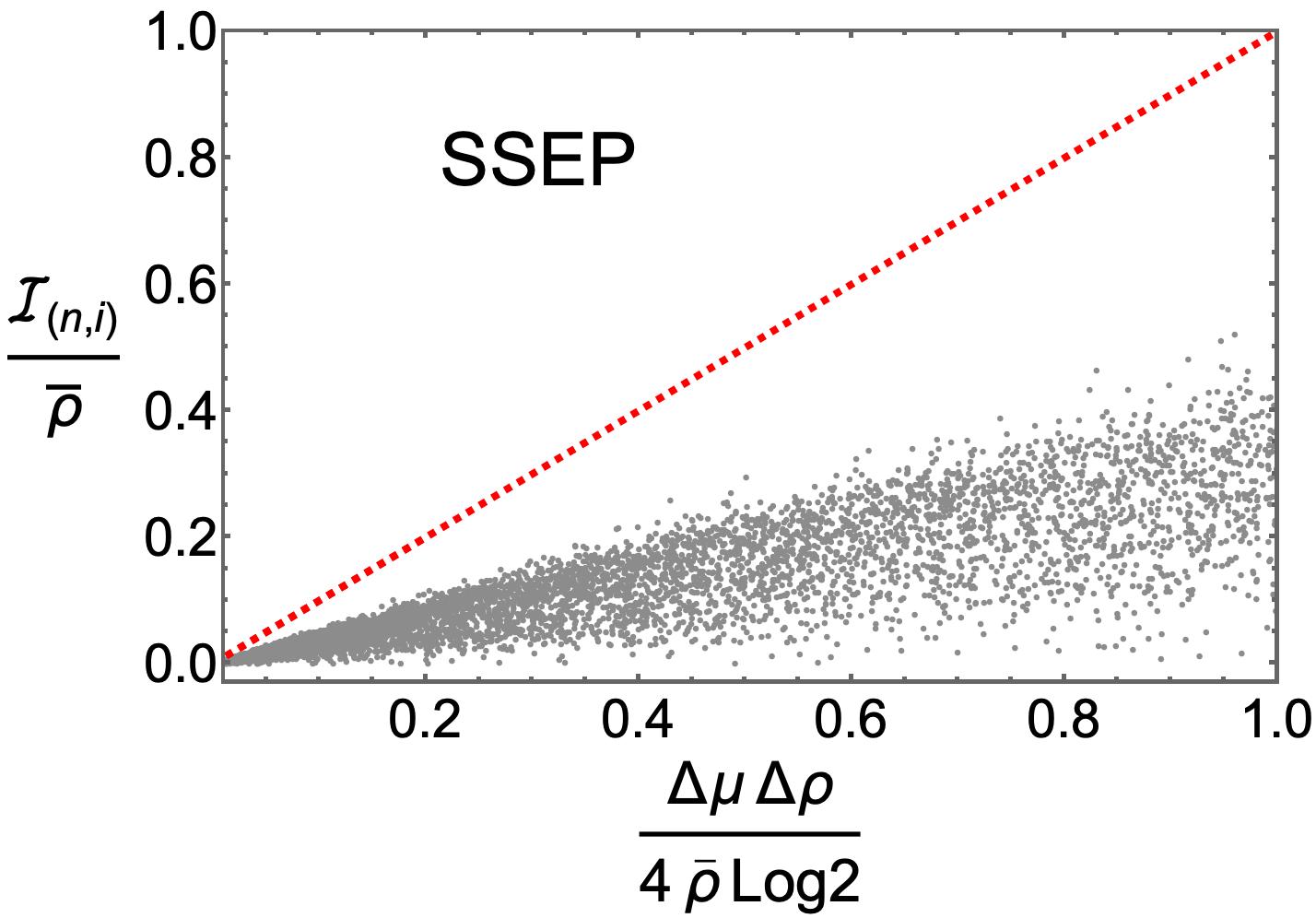}}
  \subfigure{\includegraphics[scale=0.34]{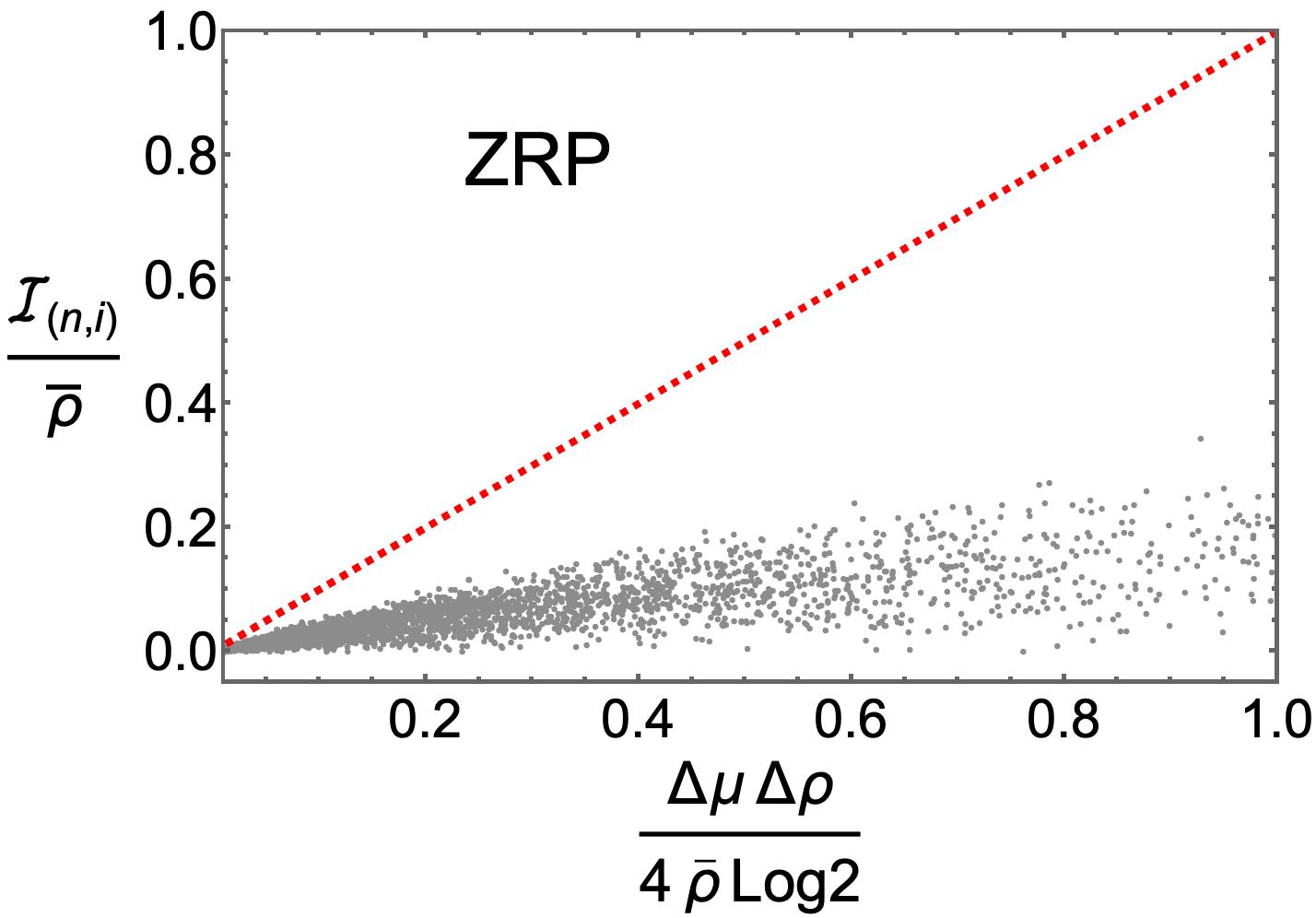}}
  \subfigure{\includegraphics[scale=0.34]{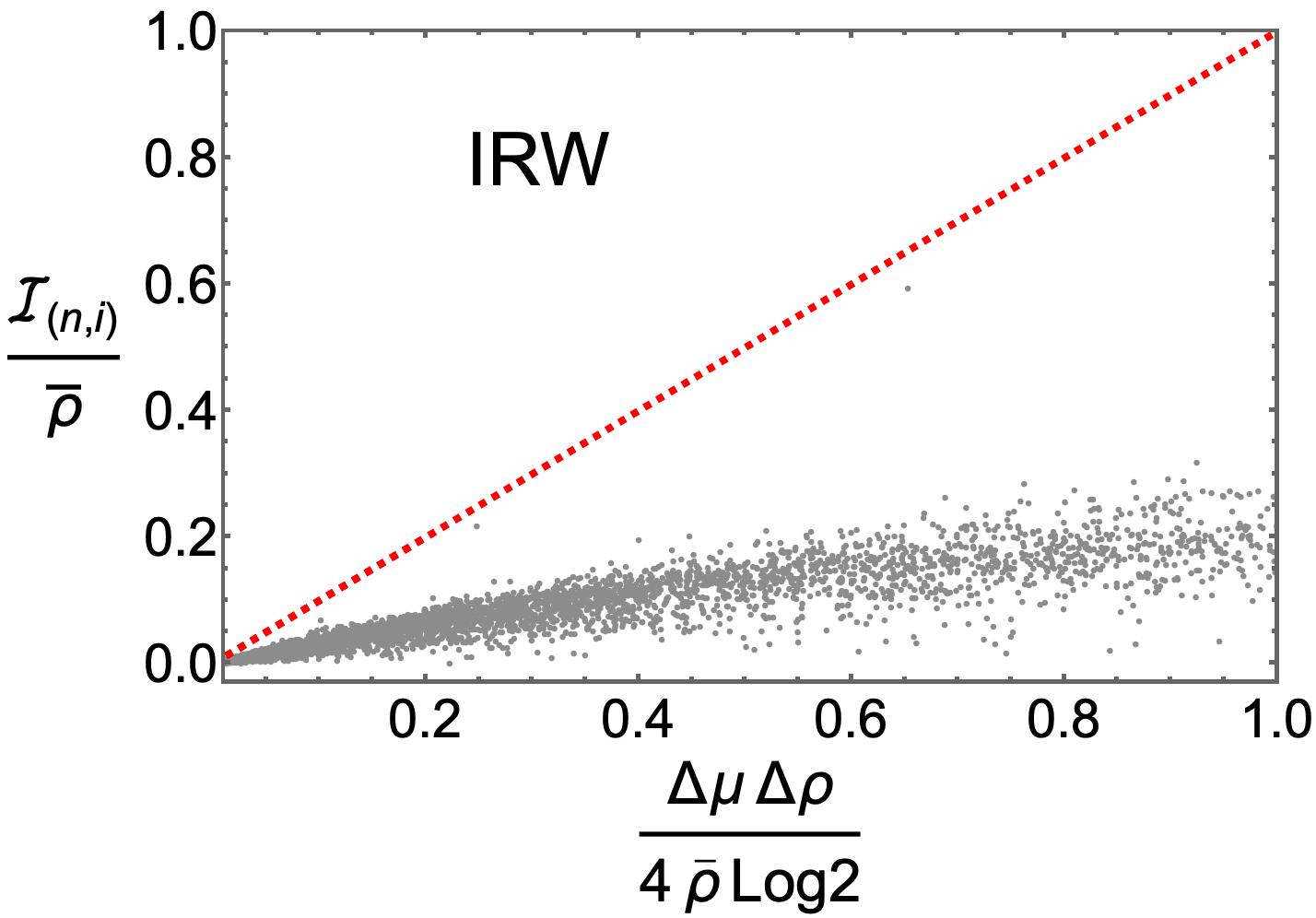}}
  \subfigure{\includegraphics[scale=0.34]{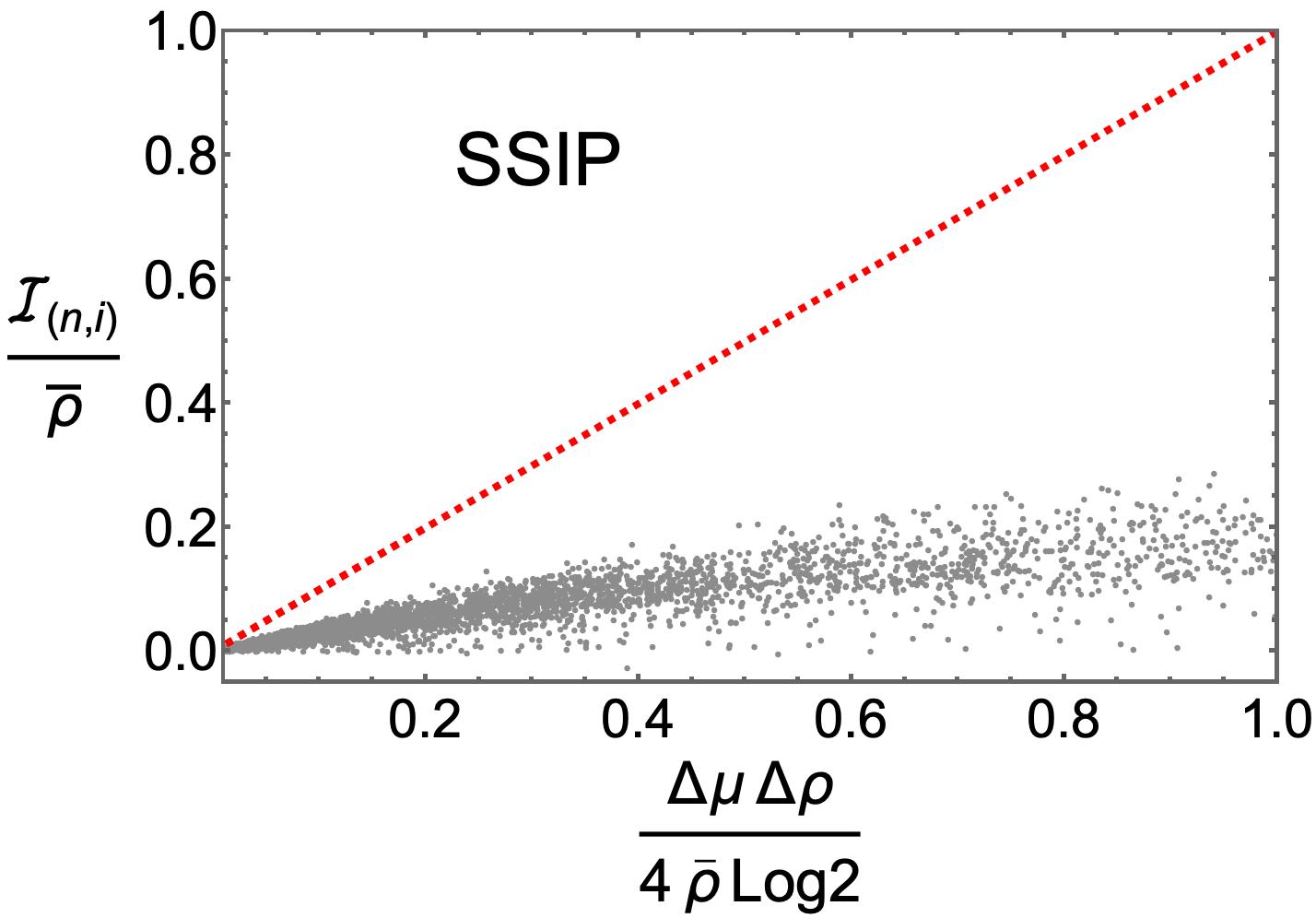}}
\centering
\caption{We have plotted $\mathcal{I} _{\left( n, i\right) } \left( \bar{\rho}, \Delta \rho \right) \Big/ \bar{\rho} $ vs $\Delta \mu \Delta \rho \Big/ 4 \bar{\rho} \ln 2$ for all four models to test the inequality~\eqref{appen-non-falt-ps-214}. For each data point, the prior probability is randomly generated by following the procedure in Eq.~\eqref{gen-priorujaho1}. We clearly see that \eqref{appen-non-falt-ps-214} is satisfied even in the strongly non-equilibrium regime.}
\label{fig-non-flat}
\end{figure}

\pss{In all the toy models studied above, we consider the case of a flat prior and numerically demonstrated the upper bound~\eqref{appen-ineqq-eq-1}. On the other hand, for a general prior, we proved that the near-equilibrium positional information satisfies a universal upper bound~\eqref{appen-non-falt-ps-214}. In this section, we will test the validity of this bound for the toy examples when the system is significantly far from the equilibrium.\\
\indent
Following the same steps as presented before, one can obtain the positional information for a general prior $P_x(x)$ to be 
\begin{align}
    \mathcal{I} _{\left( n, i\right) } \left( \bar{\rho}, \Delta \rho \right) \Big|_{\rm SSEP} = & \int _{0}^{1}dx~P_x(x) ~\left[ \rho(x) \log_2 \rho(x) + \left( 1-\rho(x)\right) \log _2 \left( 1-\rho(x)\right) \right] - \rho _{\rm ss} \log _2 \rho _{\rm ss} - (1-\rho _{\rm ss}) \log _2 (1-\rho _{\rm ss}), \nonumber \\
    \mathcal{I} _{\left( n, i\right) } \left( \bar{\rho}, \Delta \rho \right) \Big|_{\rm ZRP} = & \int _0 ^{1} dx~P_x(x) \left[ \frac{z(x) \log _2 z(x)+(1-z(x)) \log _2 (1-z(x))}{z(x)} \right] - \sum _{n=0}^{\infty} \mathcal{G}_{\rm ZRP}(n) \log _2 \mathcal{G}_{\rm ZRP}(n), \nonumber \\
   \mathcal{I} _{\left( n, i\right) } \left( \bar{\rho}, \Delta \rho \right) \Big|_{\rm IRW} = & ~ \frac{1}{\ln 2}\int _0^{1} dx~ P_x(x)~\rho(x) \left( \ln \rho(x)-1\right) - \sum _{n=0}^{\infty} \frac{\mathcal{G}_{\rm IRW}(n)}{n!} \log _2 \mathcal{G}_{\rm IRW}(n) , \nonumber \\
  \mathcal{I} _{\left( n, i\right) } \left( \bar{\rho}, \Delta \rho \right) \Big|_{\rm SSIP} = & \int _0^{1} dx~ P_x(x) \left[ \rho(x) \log _2 \rho(x) - (m+\rho(x)) \log _2 (m+\rho(x))\right] -\sum _{n=0}^{\infty} \frac{ ~\Gamma(m+n)~\mathcal{G}_{\rm SSIP}(n)}{m^{-m}~n! ~\Gamma(m)} \log _2 \mathcal{G}_{\rm SSIP}(n),  \label{appen-non-falt-ps-215}
\end{align}
where the different functions appearing in these expressions are defined as follows
\begin{align}
   & \rho(x) = \bar{\rho}+(1-2x) \Delta \rho,~~~\rho _{\rm ss} = \int _0^{1}~dx~ P_x(x)~\rho(x),~~z(x) = \frac{\bar{\rho} +\Delta \rho}{1+\bar{\rho} +\Delta \rho}-x \left(  \frac{\bar{\rho} +\Delta \rho}{1+\bar{\rho} +\Delta \rho}- \frac{\bar{\rho} -\Delta \rho}{1+\bar{\rho} -\Delta \rho} \right), \nonumber \\
   & \mathcal{G}_{\rm ZRP}(n) = \int _0^{1}dx~P_x(x)~z(x)^n (1-z(x)),~~\mathcal{G}_{\rm IRW}(n)  = \int _0^{1}dx~P_x(x)~\rho(x)^n e^{-\rho(x)},~~\mathcal{G}_{\rm SSIP}(n)  = \int _0^{1}dx~\frac{P_x(x)\rho(x)^n}{(m+\rho(x))^{m+n}}. \nonumber
\end{align}
Note that the exact expressions are for $v_c=1$ and $k_B T=1$. Furthermore, the chemical potential difference $\Delta \mu$ for these models are given in Table~\ref{Table-chempot}. Using these results, we have tested the inequality~\eqref{appen-non-falt-ps-214} in Figure~\ref{fig-non-flat} for different choices of $P_x(x)$. The prior $P_x(x)$ is sampled as
\begin{align}
    P_x(x) = \frac{\sum _{\ell=0}^{s}~ \kappa _{\ell} ~x^{\mu _{\ell}}}{\sum _{\ell=0}^{s}~\frac{ \kappa _{\ell} }{\mu _{\ell}+1}}, \label{gen-priorujaho1}
\end{align}
with $s=10$. We generate random values of $\kappa _{\ell}$ and $\mu _{\ell}$ by drawing $\kappa _{\ell}$ from a uniform distribution $[0,5]$ and $\mu _{\ell}$ from another uniform distribution $[-1,6]$, independently for each value of $\ell$. This gives us a random $P_x(x)$. Other parameters $\bar{\rho}$ and $\Delta \rho$ are obtained using the same procedure as before. Using these, we obtain one data point (gray symbol) in Figure~\ref{fig-non-flat}. This procedure is then repeated $5 \times 10^3$ times for each model. For better visibility of the plot, we have again rescaled both sides of \eqref{appen-non-falt-ps-214} by $\bar{\rho} $ and plotted $\mathcal{I} _{\left( n, i\right) } \left( \bar{\rho}, \Delta \rho \right) \Big/ \bar{\rho} $ vs $\Delta \mu \Delta \rho \Big/ 4 \bar{\rho} \ln 2$. Figure~\ref{fig-non-flat} clearly demonstrates that \eqref{appen-non-falt-ps-214} is satisfied even in the non-equilibrium regime.}

\begin{figure}[t]
  \centering
  \subfigure{\includegraphics[scale=0.6]{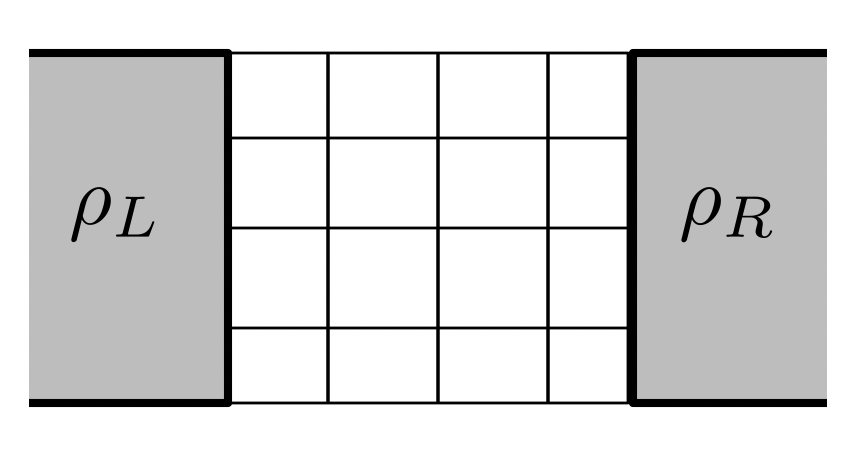}}
\centering
\caption{Higher dimensional lattice system with two particle reservoirs at densities $\rho _L$ and $\rho _R$. At other sites, a particle can jump symmetrically to any of its neighbouring sites with some rate.}
\label{fig-high}
\end{figure}
\section{Higher dimension}
\label{appen-sec-hd}
\pss{In this section, we show that the perturbative method developed in section~\ref{sppen-sec-pert} can also be extended to higher dimensional models. We consider a $d$-dimensional lattice system graph whose sites are labelled by the index $\vec{i} = (i_1,i_2, \cdots, i_d)$. As shown in Figure~\ref{fig-high}, the system is connected to two reservoirs $\rho _L$ and $\rho _R$ respectivly at $\vec{i_L} = (1,i_2,i_3 \cdots, i_d )$ and $\vec{i_R} = (N,i_2,i_3 \cdots, i_d )$ with $N$ being the separation between the two reservoirs. Such a scenario also arises in experiments where an embryo has morphogen inputs and outputs at its two poles \cite{Dubius2013}. At other sites, a particle can jump symmetrically to any of its neighbouring sites following some dynamical rule. \\
\indent
When $\rho _L$ and $\rho _R$ are well separated, i.e. $N$ is large, we anticipate the local equilibrium to hold true even in this higher dimensional set-up. Then the probability density to observe a density $g$ in a small volume $v_c$ at site $\vec{i}$ is
\begin{align}
P(g|\vec{i}) \sim  \exp \left[ - \frac{v_c}{k_B T} \Big( \mathcal{G}_{\mu_{\vec{i}}}\left( g\right) - \mathcal{G}_{\mu _{\vec{i}}}\left(  \rho_{\vec{i}}  \right) \Big) \right], \label{appen-sec-hd-1}
\end{align}
where $\rho _{\vec{i}}$ and $\mu _{\vec{i}}$ are the local average density and local chemical potential and $\mathcal{G}_{\mu_{\vec{i}}}\left( g\right)$ is given in terms of the Helmholtz free energy $a(g)$ per unit volume as
\begin{align}
\mathcal{G}_{\mu _{\vec{i}}}(g) = a(g) - \mu_{\vec{i}} g~~\text{and } \mu _i = \frac{\partial a(g)}{\partial g} \Big|_{g =  \rho _{\vec{i}}  }. \label{appen-sec-hd-2}
\end{align}
As discussed in the one-dimensional case, the precise form of $\rho _{\vec{i}} $ depends on the specific model. Yet it is possible to perform an expansion of $\rho _{\vec{i}} $ to capture the leading near-equilibrium behaviour. The first two terms in this expansion are
\begin{align}
\rho _{\vec{i}} \approx \bar{\rho} + \left(1-\frac{2i_1}{N} \right) ~\Delta \rho +\mathcal{L}_2(\vec{i}, \bar{\rho})~\Delta \rho ^2 . \label{appen-sec-hd-3}
\end{align}
To write the coefficient of the linear $\Delta \rho $ term, we use the symmetry that the density should remain invariant under the transformation $ \rho _L \leftrightarrow \rho _R$ and $(i_1, i_2, i_3, \cdots, i_d) \to (N-i_1, i_2, i_3, \cdots, i_d) $. Moreover it should be independent of $i_2, i_3, \cdots, i_d$ according to the diffusion equation.\\
\indent
Plugging this form of density in Eq.~\eqref{appen-sec-hd-1}, we can proceed exactly as in section~\ref{sppen-sec-pert} and obtain
\begin{align}
\mathcal{I} _{\left( n, \vec{i}\right) } \left( \bar{\rho} , \Delta \rho \right) \leq \frac{\Delta \mu ~v_c ~ \Delta \rho }{ 4~k_B T \ln 2}. \label{appen-sec-hd-4}
\end{align}
This is a universal bound valid for any choice of prior as long as the system is near equilibrium.}

\section{Proof of the relations in Eq.~$\eqref{FTT-appen}$}
\label{appen-relations}
\noindent
For a system in equilibrium, we stated some relations in Eq.~\eqref{FTT-appen} that give central moments of the density in terms of the derivatives of the associated free energy. Here, we present a mathematical proof of these relations. Note that the probability distribution to observe a density $n$ inside a volume $v_c$ at equilibrium is given by
\begin{align}
P_{\text{eq}}(n) = \frac{e^{- \frac{v_c}{k_B T}\left[ a(n)-a(\bar{\rho} ) - \bar{\mu}(n-\bar{\rho})\right]}}{\int dn ~e^{- \frac{v_c}{k_B T}\left[ a(n)-a(\bar{\rho}) - \bar{\mu}(n-\bar{\rho})\right]} },\label{relation-sup-eq-1}
\end{align}
where $\bar{\rho}= \langle n \rangle_{\text{eq}}$ is the average density, and $\bar{\mu}$ is the chemical potential related to $\bar{\rho}$ by
\begin{align}
\bar{\mu} = \frac{d a( \bar{\rho})}{d \bar{\rho}}.\label{relation-sup-eq-2}
\end{align}
For simplicity, we choose $k_B T=1$ and $v_c =1$. Eq.~\eqref{relation-sup-eq-1} can be further simplified as
\begin{align}
P_{\text{eq}}(n) = \frac{\exp \left[ -a(n)+\bar{\mu} n\right]}{  Z_{\text{P}}},~~~\text{with } Z_{\text{P}} = \int dn~\exp \left[ -a(n)+\bar{\mu} n\right].
\end{align}	
Any moment of the density can be computed by taking the derivative of the partition function $Z_{\text{P}}$
\begin{align}
\langle n ^{k} \rangle_{\text{eq}} = \frac{1}{Z_{\text{P}}} \frac{\partial^k Z_{\text{P}}}{\partial \bar{\mu} ^k}. \label{relation-sup-eq-3}
\end{align}	
\subsection{Variance of $n$}
Putting $k=2$ in Eq.~\eqref{relation-sup-eq-3}, we get the second moment
\begin{align}
& \langle n^2 \rangle_{\text{eq}} =  \frac{1}{Z_{\text{P}}} \frac{\partial^2 Z_{\text{P}}}{\partial \bar{\mu}^2} = \frac{\partial }{\partial \bar{\mu}} \left[ \frac{1}{Z_{\text{P}}} \frac{\partial Z_{\text{P}}}{\partial \bar{\mu} }  \right] + \left[ \frac{1}{Z_{\text{P}}} \frac{\partial Z_{\text{P}}}{\partial \bar{\mu} } \right]^2.
\end{align}	
Now identifying $\left[ \frac{1}{Z_{\text{P}}} \frac{\partial Z_{\text{P}}}{\partial \mu _L } \right] = \bar{\rho}$, one can rewrite the above expression as
\begin{align}
\sigma _2(\bar{\rho}) \equiv \langle n^2 \rangle_{\text{eq}} -\bar{\rho}^2 = \frac{d \bar{\rho}}{d \bar{\mu}}.  \label{relation-sup-eq-4}
\end{align}
To simplify further, we take $\bar{\mu}$ from Eq.~\eqref{relation-sup-eq-2} and take its derivative with $\bar{\rho}$	to get $a_2(\bar{\rho}) = \frac{d \bar{\mu}}{d \bar{\rho}}$. Plugging this in Eq.~\eqref{relation-sup-eq-4} yields
\begin{align}
\sigma _2(\bar{\rho})  = \frac{1}{a_2(\bar{\rho})}. \label{relation-sup-eq-5}
\end{align}
This relation gives us the equilibrium variance of the density in terms of the underlying free energy of the system.	
\subsection{Third central moment of $n$}
We next look at the third moment for which we put $k=3$ in Eq.~\eqref{relation-sup-eq-3}. This gives
\begin{align}
\langle n ^{3} \rangle_{\text{eq}} &= \frac{1}{Z_{\text{P}}} \frac{\partial^3 Z_{\text{P}}}{\partial \bar{\mu} ^3}= \frac{1}{Z_{\text{P}}}~ \frac{\partial}{\partial \bar{\mu}}\left[ Z_{\text{P}}~\langle n ^{2} \rangle_{\text{eq}} \right], \nonumber \\
&= \bar{\rho} ~\langle n ^{2} \rangle_{\text{eq}} + \frac{\partial \langle n ^{2} \rangle_{\text{eq}}}{\partial \bar{\mu}}. \label{relation-sup-eq-6}
\end{align}
To evaluate the derivative with $\bar{\mu}$, we use Eqs.~\eqref{relation-sup-eq-2} and \eqref{relation-sup-eq-4} as $ \frac{\partial \langle n ^{2} \rangle_{\text{eq}}}{\partial \bar{\mu}} =  \frac{\partial \langle n ^{2} \rangle_{\text{eq}}}{\partial \bar{\rho}}~\frac{d \bar{\rho}}{d \bar{\mu}} = \sigma _2(\bar{\rho})~\frac{\partial \langle n ^{2} \rangle_{\text{eq}}}{\partial \bar{\rho}}$. Furthermore, we also write $\langle n ^{2} \rangle_{\text{eq}} = \sigma _2(\bar{\rho})+\bar{\rho}^2$. The expression of $\langle n ^{3} \rangle_{\text{eq}}$ then becomes
\begin{align}
\langle n ^{3} \rangle_{\text{eq}} &= 3 \bar{\rho} ~\sigma _2(\bar{\rho}) + \bar{\rho}^3+\sigma _2(\bar{\rho}) \frac{\partial \sigma _2(\bar{\rho})}{\partial  \bar{\rho}}. \label{relation-sup-eq-7}
\end{align}
With this expression, the third central moment $\sigma _3(\bar{\rho}) = \langle (n-\bar{\rho})^3 \rangle_{\text{eq}}$ turns out to be
\begin{align}
\sigma _3(\bar{\rho}) = \sigma _2(\bar{\rho})~\frac{\partial \sigma _2(\bar{\rho})}{\partial \bar{\rho}} = -\sigma _2(\bar{\rho})^3~a_3(\bar{\rho}),
\end{align}	
where for the second equality, we have used Eq.~\eqref{relation-sup-eq-5}. Proceeding in the same, one can show that the fourth central moment is
\begin{align}
\sigma _4(\bar{\rho})  = -\sigma _2(\bar{\rho})^4 ~a_4(\bar{\rho}) + \frac{3}{\sigma _2(\bar{\rho})} \left[ \sigma _3(\bar{\rho})^2+\sigma _2(\bar{\rho})^3 \right].
\end{align}

\section{Derivation of the Helmholtz free energy}
\label{sec-appen-FE}
\ps{Here, will derive the expression for the Helmholtz free energy $a(g)$ per unit volume. For simplicity, we will focus on the zero-range process, as the mathematical procedure for other models is essentially the same. From Eq.~\eqref{FTT-appen}, we see that the second derivative of $a(\bar{\rho})$ is related to the variance of density $\sigma_2(\bar{\rho}) = \langle g^2\rangle-\bar{\rho} ^2 $ in equilibrium. Thus, we begin with the computation of the variance.\\
\indent
At equilibrium, the entire system has the same density including at the boundaries $\rho _L = \rho _R = \bar{\rho}$, and the probability to observe a density $g = n/v_c$ \pss{about a small volume $v_c$ of a lattice site} follows from Eq.~\eqref{eq-ZRP-3} to be
\begin{align}
P(g) = \frac{\left( v_c \bar{\rho} \right)^{gv_c}}{\left( 1+v_c \bar{\rho} \right)^{gv_c+1}},~~~\text{with }g = 0, \frac{1}{v_c}, \frac{2}{v_c},\frac{3}{v_c}, \cdots \label{appen-FE-eq-1}
\end{align}
Here $\bar{\rho} = \langle g \rangle$ and is related to the model parameters as
\begin{align}
v_c\bar{\rho} = \frac{\alpha _L}{\beta  _L-\alpha _L}= \frac{\alpha _R}{\beta  _R-\alpha _R}, \label{appen-FE-eq-2}
\end{align}
The density variance is
\begin{align}
\sigma_2(\bar{\rho}) = \bar{\rho} \left( \frac{1}{v_c}+\bar{\rho} \right).
\end{align}
Using Eq.~\eqref{FTT-appen}
\begin{align}
\frac{a_2(\bar{\rho})}{k_B T} = \frac{1}{\bar{\rho}(1+v_c \bar{\rho})}.
\end{align}
The integration with respect to $\bar{\rho}$ gives
\begin{align}
\frac{a_1(\bar{\rho})}{k_B T} = \ln \left( v_c \bar{\rho}\right)-\ln \left(1+ v_c \bar{\rho}\right) +\mathbb{M}, \label{appen-FE-eq-3}
\end{align}
where $\mathbb{M}$ is the integration constant. To evaluate it, we first note that when $\bar{\rho} \to \infty$, Eq.~\eqref{appen-FE-eq-2} gives $\alpha _L \to \beta _L$. This means that $\bar{\rho} \to 0$ is possible when the chemical potential $\mu _L  \sim \ln (\alpha_L / \beta _L)$ goes to zero.
Since $a_1(\bar{\rho})$ is related to the chemical potential through Eq.~\eqref{supp-general-bd-eq-2}, we must have $a_1(\bar{\rho} \to \infty) = 0$. Plugging this in Eq.~\eqref{appen-FE-eq-3} then gives $\mathbb{M} = 0$.\\
\indent
Performing one more integration 
\begin{align}
\frac{a(\bar{\rho})}{k_B T} = \bar{\rho} \ln \left(v_c \bar{\rho} \right)-\left( \frac{1}{v_c}+\bar{\rho} \right) \ln \left(1+v_c \bar{\rho} \right).
\end{align}
This gives us the Helmholtz free energy $a(\bar{\rho})$ per unit volume for the ZRP model. The same analysis can be repeated for other models as well and the results can be summarised as follows
\begin{equation}
\frac{a(\bar{\rho})}{k_B T} = \begin{cases}
 \bar{\rho} \ln \left(v_c \bar{\rho} \right) +(1/v_c-\bar{\rho}) \ln(1-v_c\bar{\rho}),~~~~~~~~\text{SSEP} \\
\bar{\rho} \ln \left(v_c \bar{\rho} \right)-\left( \frac{1}{v_c}+\bar{\rho} \right) \ln \left(1+v_c \bar{\rho} \right),~~~~~~~~\text{ZRP} \\
\bar{\rho} \ln \left( v_c\bar{\rho} \right)-\bar{\rho},~~~~~~~~~~~~~~~~~~~~~~~~~~~~~~~~~~\text{IRW}\\
\bar{\rho} \ln \left( v_c \bar{\rho} \right) -\left(\frac{m}{v_c}+ \bar{\rho} \right) \ln\left(m+v_c \bar{\rho} \right).~~~~~~~\text{SSIP}\\
\end{cases} \nonumber
\end{equation}}

%\nocite{suppb}
\bibliography{Bib_new}